\documentclass[emulateapj,epsf]{article}
\usepackage[]{emulateapj}
\def\gtsima{$\, \buildrel > \over \sim \,$}
\def\ltsima{$\, \buildrel < \over \sim \,$}
\def\simgt{\lower.5ex\hbox{\gtsima}}
\def\simlt{\lower.5ex\hbox{\ltsima}}
\def\sm{$\sim\,$}

\def\onesigma{$1\,\sigma$}
\def\nhat{\ifmmode {\hat{\bf n}}\else${\hat {\bf n}}$\fi}
\def\degs{\ifmmode^\circ\else$^\circ$\fi}
\def\kms{\ifmmode{\rm km}\,{\rm s}^{-1}\else km$\,$s$^{-1}$\fi}
\def\etal{{\sl et al.}}

\def\vev#1{{\left\langle#1\right\rangle}}

\def\bfx{{\bf x}}
\def\bfk{{\bf k}}

\def\msun{M_\odot}

\def\h1{\ifmmode h^{-1}\else$h^{-1}$\fi}
\def\kmsmpc{\kms\ {{\rm Mpc}}^{-1}}
\def\omegam{\Omega_{\rm M}}
\def\omegal{\Omega_\Lambda}

\lefthead{WILLICK}
\righthead{CONSTRAINTS ON PRIMORDIAL NONGAUSSIANITY} 
\begin{document}
\submitted{Submitted to the Astrophysical Journal}
\title{Constraints on Primordial Nongaussiantiy from the
High-Redshift Cluster MS1054--03}
\author{Jeffrey A.\ Willick\altaffilmark{1} }
\affil{Department of Physics, Stanford University, Stanford, CA 94305-4060 \\
 E-mail: jeffw@perseus.stanford.edu}
\altaffiltext{1}{Cottrell Scholar of
Research Corporation}

\begin{abstract}
The implications of the massive, X-ray selected
cluster of galaxies MS1054--03 at $z=0.83$ are
discussed in light of the hypothesis that
the primordial density fluctuations may be
nongaussian. We generalize the Press-Schechter
(PS) formalism to the nongaussian case, and 
calculate the likelihood that a cluster
as massive as MS1054 would have been found
in the Einstein Medium Sensitvity Survey (EMSS).
A flat universe ($\omegam+\omegal=1$) is assumed
and the mass fluctuation amplitude is normalized to the the present-day
cluster abundance. The
probability of finding an MS1054-like cluster then depends
only on $\omegam$ and the extent of primordial
nongaussianity. We quantify the latter by
adopting a specific functional form for the PDF, denoted
$\psi_\lambda,$ which tends to Gaussianity for $\lambda\gg 1$
but is significantly nongaussian for $\lambda \simlt 10,$ 
and show how $\lambda$ is related to the more familiar
statistic $T,$ the probability of $\ge 3\sigma$ fluctuations
for a given PDF relative to a Gaussian.
Special attention is given to a careful calculation of
the virial mass of MS1054 from the available X-ray temperature,
galaxy velocity, and weak lensing data.

We find that Gaussian initial density fluctuations are
consistent with the data on MS1054 only if $\omegam\simlt 0.2.$
For $\omegam\ge 0.25$ a significant degree of
nongaussianity is required, unless
the mass of MS1054 has been substantially overestimated
by X-ray and weak lensing data. The required amount of nongaussianity
is a rapidly increasing function of $\omegam$ for $0.25 \le \omegam \le 0.45,$
with $\lambda \le 1$ ($T \simgt 7$) at the upper end of this range.
For a fiducial $\omegam=0.3,$ $\omegal=0.7$ universe, favored
by several lines of evidence (Wang \etal\ 1999),
we obtain an upper limit $\lambda \le 10,$
corresponding to a $T\ge 3.$ 
This finding is consistent with the conclusions
of Koyama, Soda, \& Taruya (1999), who applied
the generalized PS formalism to low ($z\simlt 0.1$)
and intermediate ($z\simlt 0.6$) redshift cluster data sets.
\end{abstract}

\section{Introduction}
\label{sec:intro}
A working hypothesis in most approaches to 
cosmological structure formation
is that the density fluctuation field $\delta(\bfx)$ is Gaussian
at early times. This assumption follows
naturally from the idea that the primordial fluctuations
may be described as superpositions of statistically independent
Fourier modes,
\begin{equation}
\delta(\bfx) = \sum \delta_\bfk e^{i\bfk\cdot\bfx}\,.
\label{eq:fourier}
\end{equation}
Such modes are 
generic predictions of inflation, 
in which quantum fluctuations
are exponentially enlarged to become the 
classical density fluctuations. 
Whatever the form of
the power spectrum $P(k) = |\delta_\bfk|^2,$
the independence of the Fourier modes
guarantees, by the central limit theorem, that the 
real-space density fluctuations are Gaussian.

Although the Gaussian hypothesis is well-motivated,
several lines of argument suggest 
that it merits further scrutiny.
On the theoretical side,
Peebles (1999a,b) 
has presented a model in which nongaussianity
emerges naturally in a modified inflationary scenario.
The key ingredient is that the scalar field which drives
inflation, $\psi,$ couples quadratically to a massive scalar
field $\phi$ which, at the end of inflation, produces the cold dark matter
(CDM). The 
CDM density field is then given by $\rho_{CDM}(\bfx)
=\mu^2 \phi^2(\bfx)/2,$ where $\phi(\bfx)$ is a Gaussian
random field. The fluctuations in the CDM
component, $\delta(\bfx) \propto \rho_{CDM}(\bfx)-\vev{\rho_{CDM}},$
are thus distributed like a $\chi^2$ function shifted to
have vanishing mean. 
Such a distribution is strongly nongaussian, with significant
skew and excess kurtosis. (Other inflationary models
which lead to nongaussian density fluctuations have
recently been described by Salopek [1999] and by
Martin, Riazuelo, \& Sakellariadou [1999].)  

Peebles' model has been
extended by White (1998) and by Koyama, Soda, \& Taruya (1999),
who have considered models in which the CDM 
is produced by
a finite number, $m,$ of quadratically coupled fields. 
The resulting density fluctuations would then
have a $\chi^2_m$ distribution (again, shifted to have mean zero),
which approaches Gaussianity
in the limit of large $m.$ In this way one can generate
primordial fluctuations
with any desired amount of nongaussianity. However, the required
number of quadratically coupled fields must be quite large ($\simgt100$) 
if the wanted amount of nongaussianity is small. Such a large
number of inflationary fields arguably makes this model contrived. 

On the observational side, hints of nongaussianity have emerged, first,
from Cosmic Background Radiation (CBR) anisotropy maps. 
Ferreira \etal\ (1998, 1999; see also Magueijo 1999) 
analyzed the full-sky COBE-DMR maps (Bennett \etal\ 1996),
finding evidence of nongaussianity on large ($\simgt 10\degs$) angular scales
using a statistic known as the normalized
bispectrum estimator. This finding was confirmed
by two independent analyses using alternative statistical
techniques (Novikov, Feldman, \& Shandarin, 1998;
Pando, Valls-Gabaud, \& Fang, 1998).
These studies appear to rule out
Gaussianity of the CBR anisotropies at about the
95\% confidence level. A curious feature
of these COBE-DMR results is that the nongaussian signal 
goes away if the
north galactic cap is excluded from the analysis; this is
a surprising property, one that argues for caution in
ascribing reality to these findings (see Bromley \& Tegmark 1999
for a detailed treatment of this issue).
In an unrelated study of 
CBR anisotropy detections on degree
scales, Gazta\~naga, Fosalba, \& Elizalde 
(1998) found evidence for nongaussianity
based on the large variance among the reported
anisotropy amplitudes. This finding, like
that of Ferriera \etal, 
should be considered preliminary, as it is based
on an intercomparison of very different data sets 
and is strongly dependent on the accuracy of the reported errors.
Nonetheless, these results from CBR anisotropy
data call into question the hypothesis of
primordial gaussianity.

The second line of observational evidence comes from the
abundance and clustering of rich clusters of galaxies.
As the most massive virialized systems in the universe,
rich clusters are diagnostic of the background cosmological parameters
in a number of ways (cf.\ Bahcall 1999
for a recent review). In particular, their number density
and correlation length, as
a function of mass and redshift, can be predicted from
the Press-Schechter (1974, hereafter PS) formalism,
summarized in \S~\ref{sec:PS} below. The original PS formalism assumes
primordial Gaussianity, but this assumption is readily
relaxed, leading to a generalized PS approach (\S~2.3).
Robinson, Gawiser, \& Silk (1998) applied the
generalized PS formalism to  low-redshift
($z\simlt 0.1$) cluster abundance and correlation data.
They found the data to be
consistent with the Gaussian hypothesis
in an $\omegam=1$ universe.\footnote{Here and throughout this
paper, the present value of the mass density parameter is denoted
$\omegam,$ while the density parameter associated
with a cosmological constant $\Lambda$ (or, equivalently,
vacuum energy density) is denoted $\omegal.$  
The Hubble constant is written as $H_0=100\,h\ \kmsmpc.$}
For lower values of the density parameter, however,
Robinson \etal\ found that
a substantial degree of primordial nongaussianity is
required to reconcile the number density of massive
clusters with their observed correlation length. Koyama,
Soda, \& Taruya (1999) further narrowed these constraints
by applying the generalized PS formalism to intermediate-redshift
($z\simeq 0.5$--0.6) cluster abundance data, as well as to the
low redshift data sets studied by Robinson \etal\ They
found that $\omegam>0.5$ was ruled out by the combined data sets
regardless of the nature of the initial density fluctuations,
and that Gaussian fluctuations were ruled out for $\omegam\leq 0.5.$
We defer a fuller discussion of their results to
the final section of this paper, after introducing the necessary
terminology in \S~\ref{sec:nongauss}.

The studies cited above provide
evidence that the hypothesis of Gaussian density fluctuations
is violated at some level. The purpose of this paper is to extend
the analyses of Robinson \etal\ (1998) and Koyama \etal\ (1999)
by applying the generalized PS approach to a single high-redshift
cluster, MS1054--03, which at $z=0.83$ is one of the highest redshift
clusters known. It is, moreover, the most extensively studied
of the known high-redshift clusters; its mass has been accurately
estimated by the techniques of weak gravitational lensing
(Luppino \& Kaiser 1997), X-ray temperature analysis (Donahue \etal\ 1998),
and galaxy velocity dispersion (Tran \etal\ 1999). These studies
all found that MS1054 is an unusually massive cluster, with
a virial mass $\simgt 10^{15}\,\h1\msun.$ Its
high redshift, large and accurately determined mass,
and its having been selected from
the Einstein Medium Sensitivity Survey (Henry \etal\ 1992; EMSS),
whose selection criteria are very well understood, combine to
make MS1054 especially well suited for the present study.
A previous analysis of MS1054 (Bahcall \& Fan 1998) noted
these features, and derived important cosmological constraints,
but did so under the assumption of primordial Gaussianity. 
We will extend their results by including nongaussianity
as an additional degree of freedom.

This paper is also meant to serve two additional purposes.
First, we will introduce (\S~\ref{sec:nongauss}) a new mathematical
description of nongaussian density fluctuations, in the form of
a probability distribution function (PDF) we refer to as the
$\psi_\lambda$-distribution. We will discuss the relationship
between the $\psi_\lambda$ and $\chi^2_m$ distributions, 
and argue that the former is perferable to the extent that
a fully generic, model-independent parameterization of
nongaussianity is desired. Second, we will present 
(\S~\ref{sec:clmass}) a detailed
discussion of the proper determination of cluster virial masses
from X-ray temperature, galaxy 
velocity dispersion, and weak lensing data sets, and apply the results
to MS1054. This discussion will, it is hoped, be useful in
clarifying the ways in which virial masses are dependent
on cosmological parameters and, more subtly, on cluster mass
models, and thus inform future studies based on larger data sets.

The outline of this paper is as follows. In \S~\ref{sec:PS}, we
describe the Gaussian and generalized PS formalisms. In
\S~\ref{sec:nongauss}, we introduce the $\psi_\lambda$ distribution.
In \S~\ref{sec:clmass} we discuss the determination of cluster
virial masses.
In \S~\ref{sec:ms1054} we use published data to estimate
the virial mass of MS1054.
In \S~\ref{sec:results} we apply the generalized PS formalism
to MS1054 and derive constraints on nongaussianity as a function
of $\omegam.$ Finally, in \S~\ref{sec:disc},
we further discuss and summarize the main results of the paper.

\section{The Press-Schechter Formalism and its Extension to 
the Nongaussian Case}
\label{sec:PS}

Under the assumption of primordial Gaussianity,
the PS formula for the comoving number density of virialized objects
of mass $M$ is
\begin{equation}
n(M,z) = \sqrt{\frac{2}{\pi}}\,\frac{\overline\rho}{M^2}\,
\frac{\delta_c(z)}{\sigma_M}\left|\frac{d\ln \sigma_M}{d\ln M}\right| 
\,e^{-\delta_c(z)^2/2\sigma_M^2}\,,
\label{eq:PS}
\end{equation}
where $\sigma_M$ is the rms mass fluctuation on a mass scale $M,$ 
$\overline\rho\equiv\omegam\rho_{crit}$ is the comoving mean
mass density, and $\delta_c(z)$ is the critical density
for collapse at redshift $z.$ Because $\sigma_M$ is by
definition the rms density fluctuation linearly
extrapolated to the present,
$\delta_c(z)$ is similarly normalized to the present, 
\begin{equation}
\delta_c(z;\omegam,\omegal) = \delta_0(z) \frac{D(z=0;\omegam,\omegal)}
{D(z;\omegam,\omegal)}\,.
\label{eq:deltacz}
\end{equation} 
The linear fluctuation growth factor is given by
\begin{equation}
D(z;\omegam,\omegal) = \frac{5}{2} \omegam E(z) \int_z^{\infty}
\frac{1+z'}{E(z')^3} dz'\,,
\label{eq:dz}
\end{equation}
where $E(z)=\sqrt{\omegam(1+z)^3+\Omega_R(1+z)^2+\omegal},$
$\omegam+\Omega_R+\omegal=1,$
and we are following Peebles' (1993) notation.
Throughout this paper we assume a flat universe, $\Omega_R=0.$

The quantity $\delta_0(z)$ in equation~\ref{eq:deltacz} is the
famous factor $3(12\pi)^{2/3}/20=1.686$ 
for an Einstein de-Sitter universe. It has
a weak dependence on cosmological parameters and redshift. 
In this paper, we use the forms of $\delta_0(z)$
derived by Kitayama \& Suto (1996). 

\subsection{Calculation of $\sigma_M$}

The rms mass fluctuation $\sigma_M$ in equation~\ref{eq:PS}
is computed from the linear
power spectrum $P(k)$ as follows:
\begin{equation}
\sigma_M^2 = \int_0^\infty \frac{dk}{k} \Delta^2(k)\,W^2(kR)\,,
\label{eq:sigmr}
\end{equation}
where $\Delta^2(k)\equiv k^3 P(k)/2\pi^2$ is the mass variance
per logarithmic wavenumber interval, $W(kR)$ is the
Fourier transform of the window function defining the
mass scale $M,$ which we take to be the usual top-hat,
and $R$ is the radius of a sphere which contains
mass $M$ in an unperturbed universe.
In this paper we assume a CDM-dominated universe, 
in which the
power spectrum is well-approximated by
\begin{equation}
\Delta^2(k) = \delta_H^2(\omegam,\omegal) 
\left(\frac{ck}{H_0}\right)^{3+n}\,T^2(k/\Gamma)
\label{eq:defdelta}
\end{equation}
(Bunn \& White 1997), where $\delta_H$ is the
rms overdensity at horizon-crossing, $n$ is the
primordial spectral index, and $T(q)$ is the
CDM transfer function, for which we adopt
the analytic approximation of Bardeen \etal\ (1986). 
The parameter $\Gamma$ determines
the position of the power-spectrum ``turnover'' in $k$-space.
For CDM it is given by $\Gamma \approx \omegam h.$ 
For the calculations of this paper, the precise value
of $\Gamma$ is relatively unimportant,
and we adopt the value $\Gamma = 0.20,$ consistent
with observations of large-scale structure data (e.g.,
Liddle \etal\ 1996) and with the CDM expectation
for reasonable values of the cosmological parameters. We also  assume
$n=1,$ consistent with estimates derived from
the large-scale CBR anisotropies observed by
COBE (Gorski \etal\ 1996). Changing our adopted values
of $\Gamma$ and $n$ by $\simlt 20\%,$ roughly their
allowed ranges, would have little effect on the main conclusions
of this paper.

\subsection{Conversion to useful units}

To make practical calculations,
we reexpress the PS formula in terms of suitably scaled variables.
We first define
a dimensionless mass $m$ by
\begin{equation}
M = \frac{4\pi}{3} r_8^3 \rho_{crit}\, m  = 5.95\times 10^{14} m\, \h1\ M_\odot\,,
\label{eq:defm}
\end{equation}
where $r_8\equiv 8\,\h1$ Mpc and $\rho_{crit}=3H_0^2/8\pi G$ 
is the critical density. 
For rich clusters, $m$ is of order unity.
Let $n(m) dm$ be the comoving number density of clusters
with dimensionless masses in the range $(m,m+dm).$ Then
\[ n(m,z) = n(M,z) \frac{dM}{dm} \] 
\begin{equation}
= \sqrt{\frac{2}{\pi}} \frac{3\,\omegam}
{4\pi\,r_8^3\,m^2} \frac{\delta_c(z)}{\sigma_M} 
\left|\frac{d\ln \sigma_M}{d\ln M}\right| e^{-\delta_c(z)^2/2\sigma_M^2}\,,
\label{eq:ps1}
\end{equation}
where we have 
substituted the definition of $m$ into equation~1.
Furthermore, as noted above, we compute $\sigma_M$
not as a function of mass but of a length scale $R_m$ defined by
$M=\frac{4}{3}\pi R_m^3 \omegam \rho_{crit},$
or equivalently,
\begin{equation}
R_m = r_8\left(\frac{m}{\omegam}\right)^{1/3}\,.
\label{eq:defrm}
\end{equation}
Thus, $d\ln\sigma_M/d\ln M=1/3\times d\ln\sigma_M/d\ln R.$
Substituting this into equation~\ref{eq:ps1} and evaluating numerical
factors yields
\begin{equation}
n(m,z) = n_0\,\frac{\omegam}{m^2} \left|\frac{d\ln\sigma_M}
{d\ln R}\right|_{R_m}\!\nu_m \phi(\nu_m)\,, 
\label{eq:ps2}
\end{equation}
where $n_0=3.11\times 10^{-4}(\h1{\rm Mpc})^{-3},$ 
$\phi(x)$ is a Gaussian of zero mean 
and unit variance, and
$\nu_m(z) \equiv \delta_c(z;\omegam)/\sigma_M(R_m).$ 
Equation~\ref{eq:ps2}
gives the comoving number density per unit dimensionless mass;
note that most of the $\omegam$-dependence of
this expression resides in the factor $\nu_m,$ not in the
$\omegam$ out in front.
Finally, the directly observed quantity is 
$N(\ge\!m,z),$ the comoving
number density of all clusters of mass $\ge m$ at redshift $z,$ 
given by
\begin{equation}
N(\ge\!m,z) = n_0\, \omegam\!\int_m^\infty \frac{dm'}{m'^2}
\left|\frac{d\ln\sigma_M}{d\ln R}\right|_{R_{m'}}\!\nu_{m'} \phi(\nu_{m'})\,.
\label{eq:numden}
\end{equation}

\subsection{Extension to Nongaussian Fluctuations}

The above formulae assumed a Gaussian PDF.
We now relax this assumption, and
assume only that the PDF, $P(\delta|M),$ is such that $\vev{\delta}=0$
and that $\vev{\delta^2}=\sigma_M^2.$ It is useful to express the PDF
in terms of a dimensionless function $\psi(x)$ as follows:
\begin{equation}
P(\delta|M) = \sigma_M^{-1}\,\psi\left(\frac{\delta}{\sigma_M}\right)\,,
\label{eq:pgen}
\end{equation}
where $\psi(x)$ satisfies the conditions
$\int \psi(x)\,dx=1,$
$\int x\, \psi(x)\,dx =0,$ and
$\int x^2\,\psi(x)\,dx =1\,.$

If one now retraces the usual steps leading
to the PS abundance formula, equation~\ref{eq:PS},
but using the PDF given by equation~\ref{eq:pgen} in
place of the usual Gaussian, one arrives at the expression
\begin{equation}
n(M,z) = \frac{2 f_\psi\,\overline\rho}{M^2}\,
\frac{\delta_c(z)}{\sigma_M}\left|\frac{d\ln \sigma_M}{d\ln M}\right|
\psi\left(\frac{\delta_c(z)}{\sigma_M}\right)\,.
\label{eq:psgen}
\end{equation}
The quantity $f_\psi$ is given by 
$\left(2 \int_0^\infty \psi(x)\,dx\right)^{-1}$;
it corrects for underdense regions that are incorporated into
virialized structures according to the standard PS ansatz.
Equation~\ref{eq:psgen} is our generalization of the PS formalism for non-Gaussian 
density perturbations. Setting $\psi(x)$ to a Gaussian
yields equation~1, as may readily be verified.
Adopting dimensionless mass units and repeating
the steps of \S~2.2, we obtain for $N(\ge\!m,z)$
the result
\begin{equation}
N(\ge\!m,z) = f_\psi n_0\, \omegam\!\int_m^\infty \frac{dm'}{m'^2}
\left|\frac{d\ln\sigma_M}{d\ln R}\right|_{R_{m'}}\!\nu_{m'} \psi(\nu_{m'})\,,
\label{eq:numden_gen}
\end{equation}
where the meaning of $n_0$ and $\nu_m$ are the same as above.
The difference between the standard and generalized
PS abundance predictions thus consists simply in replacing
the Gaussian PDF $\phi$ by the function $\psi,$
apart for the factor $f_\psi,$ which as shown below
is $\sim 1$ for cases of interest.

\section{Description of NonGaussianity}
\label{sec:nongauss}
We wish to consider non-Gaussian fluctuations generically,
i.e., to construct a PDF meeting the above conditions
but which is not tied to a particular model. Moreover, we want
our distribution to contain a parameter which
quantifies the degree of nongaussianity. 
As noted above, a possible
choice is the $\chi^2_m$ distribution, a generalization
of the Peebles (1999a,b) model proposed by
White (1998) and by Koyama \etal\ (1999). Here
we suggest an alternative parameterization.
Our reasons for doing so are twofold: first,
the $\chi^2_m$ distribution is associated
with a particular physical model, whereas
our preference is to avoid such association
at this preliminary stage in our understanding
of the primordial fluctuations; and second,
the $\chi^2_m$ model approaches gaussianity only
in the limit of very large $m,$ a fact which makes
the physical interpretation difficult to sustain
if the observationally
required degree of nongaussianity is slight. 

Our model is based on a modified form of the Poisson distribution.
Consider, first, an integer random variable $n$ 
that is Poisson-distributed with expectation
value $\lambda.$ The probability that $n$ takes on a particular integer
value $m$ is 
\begin{equation}
P(n=m) = \frac{\lambda^m}{m!} e^{-\lambda}\,.
\label{eq:poisson}
\end{equation}
The rms deviation of $n$ is $\sqrt{\lambda}.$ 
We now imagine that $n$ is continuous rather than discrete,
and define a related random variable 
$x\equiv (n-\lambda)/\sqrt{\lambda}.$  
The probability
distribution of $x,$ which we refer to as $\psi_\lambda(x),$
is a modified form of the Poisson distribution,
but shifted
and scaled so that it has mean zero and unit variance.
We obtain an explicit representation of $\psi_\lambda(x)$ by
letting $m=\sqrt{\lambda} x + \lambda$
in equation~\ref{eq:poisson}, and then multiplying by $\sqrt{\lambda}$ to
renormalize the distribution. In addition,
the factorial function in the denominator, which is defined
only for integer values of its argument, must be replaced by its
appropriate generalization for continuous variables, the $\Gamma$-function.
This yields
\begin{equation}
\psi_\lambda (x) = \frac{\lambda^{\sqrt{\lambda}x+\lambda+\frac{1}{2}}\,
e^{-\lambda}}
{\Gamma(\sqrt{\lambda}x+\lambda+1)}  \,.
\label{eq:defpsi}
\end{equation}
This function is defined for $x>-(\sqrt{\lambda}+1/\sqrt{\lambda}).$ 
As we now show,
$\psi_\lambda(x)$ is a suitable representation
of quantifiable, generic departures from Gaussianity.\footnote{We
note that equation~\ref{eq:defpsi} does
not guarantee that $\psi_\lambda(x)$ will have the key properties  
we seek: normalization, vanishing mean, and unit
variance.  For small $\lambda,$ 
the transition from integer to continuous arguments
does not preserve these properties, which were inherent
in the parent Poisson distribution. By direct integration we
have found that departures from these
properties are completely negligible for
$\lambda > 10.$ For $\lambda < 10,$ we correct 
$\psi_\lambda(x)$ using formulae derived by fitting
the deviations from normalization, vanishing mean,
and unit variance. However, these corrections are
extremely small for the cases of interest, so that
equation~\ref{eq:defpsi} is fully adequate for our purposes.}

\vbox{%
\begin{center}
\leavevmode
\hbox{%
\epsfxsize=8.9cm
\epsffile{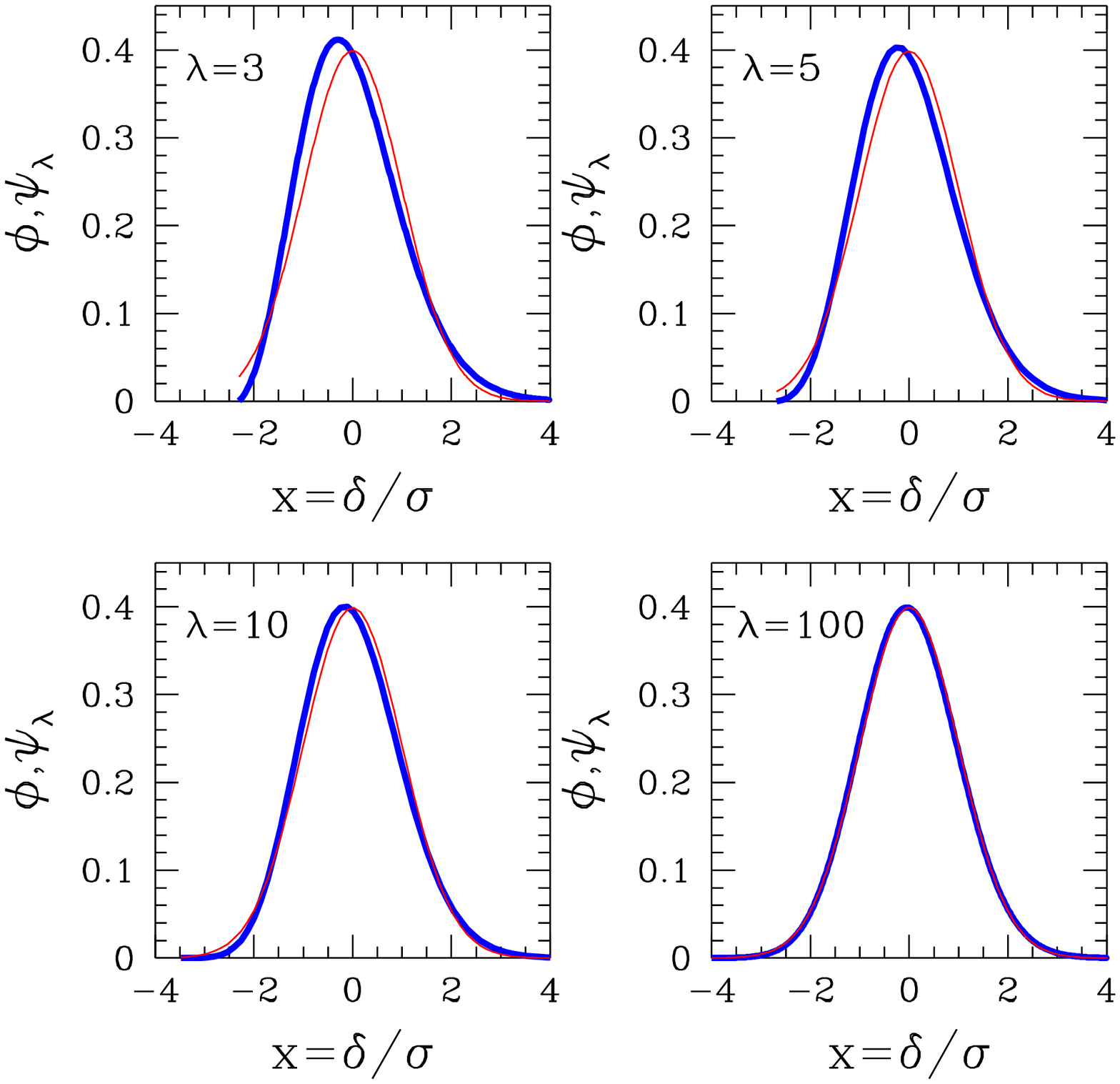}}
\begin{small}
\figcaption{%
Comparison of a Gaussian, $\phi$ (thin red curve)
with the $\psi_\lambda$ distribution (heavy blue curve), 
equation~\ref{eq:defpsi}, for four values of
the parameter $\lambda.$ Note the progression of $\psi_\lambda$
toward Gaussianity for $\lambda\gg 1.$
\label{fig:phi_psi_1}}
\end{small}
\end{center}}

Figure~\ref{fig:phi_psi_1} compares the $\psi_\lambda$
distribution with a Gaussian for four values
of the parameter $\lambda.$ For $\lambda=3,$ the
differences between the two are readily apparent;
in particular, one sees the significant skewness
of the $\psi_\lambda$ distribution. For $\lambda=5,$
analogous differences can be seen, but they are
noticeably smaller, and for $\lambda=10$ the differences
are very small. For $\lambda=100,$ the $\psi_\lambda$
is indistinguishable, on this plot, from a Gaussian.

The differences between $\psi_\lambda$ and
a Gaussian, $\phi,$ become more apparent, even for $\lambda\simgt 10,$
if we examine their behavior for large values of
$\delta/\sigma.$ This is shown in Figure~\ref{fig:phi_psi_2}
for the same four values of $\lambda$ as in the previous figure.
We see that for $\lambda \simlt 10,$ $\psi_\lambda(x)$
can exceed $\phi(x)$ by 2--3 or more orders of magnitude
for $\delta/\sigma \simgt 4.$ As noted in \S~\ref{sec:PS},
this same factor enters directly into the PS-predicted cluster abundance. 
Thus, we expect that if the PDF is described by the $\psi_\lambda$
distribution, massive clusters, representing rare peaks
in the initial density field, will be far more abundant
than they will in the Gaussian case, provided $\lambda$
is not too large.
\vbox{%
\begin{center}
\leavevmode
\hbox{%
\epsfxsize=8.9cm
\epsffile{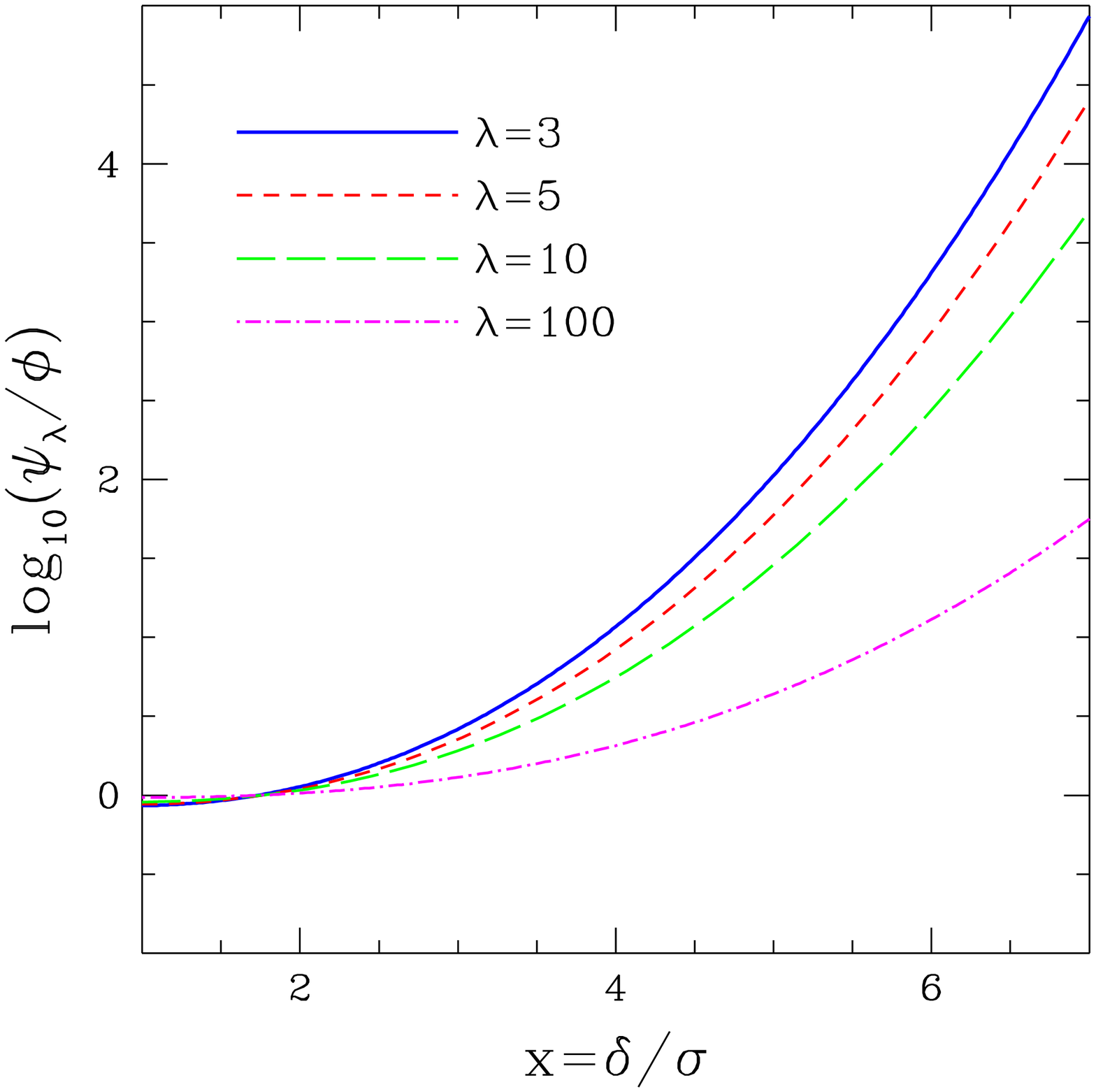}}
\begin{small}
\figcaption{%
The ratio of $\psi_\lambda$ to a Gaussian, $\phi,$
plotted for values of the argument $x \ge 1.$ The
plotted ratio shows that ``rare events,'' 
$\delta/\sigma \simgt 3,$ are considerably
more likely if the distribution of overdensities
is described by the $\psi_\lambda$ function
for $\lambda \simlt 10,$ than they are
in the Gaussian case.
\label{fig:phi_psi_2}}
\end{small}
\end{center}}

\vbox{%
\begin{center}
\leavevmode
\hbox{%
\epsfxsize=8.9cm
\epsffile{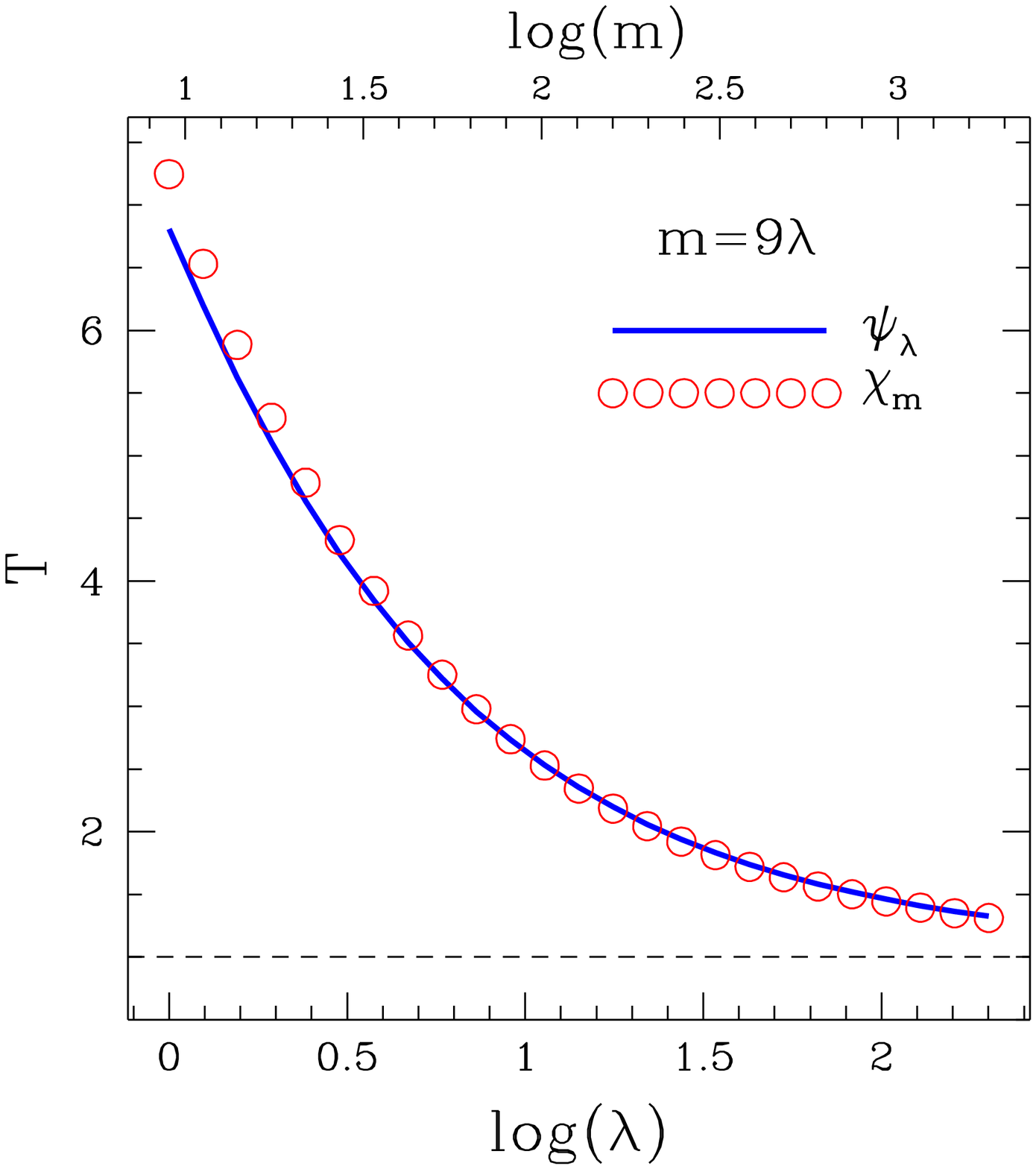}}
\begin{small}
\figcaption{%
The T statistics of the $\psi_\lambda$ (solid line)
and $\chi^2_m$ distributions (open circles) plotted
versus their respective parameters, $\lambda$ and $m.$
The axes are scaled such that $m=9\lambda.$ The
close agreement of the T statistics for the two
distributions reveals their strong similarity 
in terms of the likelihood of rare ($\simgt 3\sigma$)
events, if one chooses $m=9\lambda.$
\label{fig:Tpsi}}
\end{small}
\end{center}}

The factor $f_\psi$ that appears
in equation~\ref{eq:psgen} may be calculated numerically
for the $\psi_\lambda$ distribution. We have done so
for $2 \leq \lambda \leq 300.$ Within this range the
numerical results are very well approximated by the
expression
\begin{equation}
f_\psi(\lambda) = 1 + 0.149 \lambda^{-0.528}\,.
\label{eq:fcorr}
\end{equation}
Note that $f_\psi$ differs from unity by less than 10\%
for $\lambda\simgt 2.$ 

\subsection{Comparison with the $\chi^2_m$ distribution}
The model introduced by
White (1998) and by Koyama, Soda,
\& Taruya (1999) postulates that the CDM
density field is given by
\begin{equation}
\rho_{CDM}(\bfx) = \frac{\mu}{2}\sum^m_{i=1} \phi_i^2(\bfx) 
\label{eq:cdmchi}
\end{equation}
where the $\phi_i(\bfx)$ are statistically independent
Gaussian fields.
The overdensity $\nu=\delta/\sigma$ in this model
is distributed like a $\chi^2$ variable with $m$
degrees of freedom, shifted to have vanishing
mean and unit variance. Explicitly,
\begin{equation}
P(\nu) d\nu = \frac{\left(1+\sqrt{\frac{2}{m}}\nu\right)^{
m/2-1}}{\left(\frac{2}{m}\right)^{(m-1)/2}
\Gamma\left(\frac{m}{2}\right)}
\exp\left(-\frac{m}{2}-\sqrt{\frac{m}{2}}\nu\right) d\nu\,,
\label{eq:chi2m}
\end{equation}
a distribution we henceforth label $\chi^2_m.$

Koyama, Soda, \& Taruya (1999) (see also Robinson, Gawiser,
\& Silk 1998) have advocated 
quantifying ``rare event'' nongaussianity 
in terms of a ``T-statistic'' defined by'
\begin{equation}
T = \frac{\sqrt{2\pi} \int_3^\infty P(\nu) d\nu}{\int_3^\infty
e^{-\nu^2/2} d\nu}\,,
\label{eq:defT}
\end{equation}
the likelihood relative to Gaussian of $3\,\sigma$
or rarer events. In terms of this statistic
the relationship between the $\psi_\lambda$ 
and $\chi^2_m$ distributions becomes particularly
clear. In Figure~\ref{fig:Tpsi}, the T statistics
for the two distributions are plotted versus their
respective parameters, with the axes scaled
such that $m=9\lambda.$ With
this scaling, the two distributions have T statistics
which agree to within a few percent for all $\lambda.$  
This shows that in terms of the likelihood of rare
events, the $\psi_\lambda$ and $\chi^2_m$ distributions
are extremely similar for $m=9\lambda.$ 
It is also interesting to note that, despite the 
near-indistinguishability of $\psi_\lambda$ from
a Gaussian for $\lambda\simgt 100$ (Figure 1),
the two distributions approach the Gaussian value
$T=1$ extremely slowly as $\lambda\rightarrow\infty.$
Indeed, one requires $m \simeq 200$ even to
achieve $T=2,$ twice the Gaussian value. It is for
this reason that the $\chi^2_m$ model may not
be suitable for describing mild Gaussianity---at least
if one were to take its physical basis seriously---because
the required number of primordial Gaussian fields is
excessively large.

\section{On the Determination of Cluster Virial Masses}
\label{sec:clmass}
It is essential when applying PS abundance estimates
that one use the rigorous definition
of virial mass. Specifically, the PS formulae
apply to the mass $M_V$ interior to a radius
$r_V$ such that $3 M_V/4\pi r_V^3 = \Delta_V(z,\omegam,\omegal)
\overline\rho (z),$ where $\Delta_V(z,\omegam,\omegal)$
is a cosmology- and redshift-dependent overdensity factor.
In an Einstein-de Sitter universe, $\Delta_V\simeq 178$
at all redshifts; for $\omegam<1,$ $\Delta_V$ is larger
than this value, and increases with decreasing redshift.
Kitayama \& Suto (1996) have derived analytic
approximations for $\Delta_V(z)$ as a function of
$\omegam$ for flat and open cosmologies, and we
use their formulae in this paper. 

The definition of virial mass above means
that one cannot compute $M_V$ from 
observational data, such as velocity dispersion,
X-ray temperature, or weak lensing, without specifying
both a cosmology ($\omegam$ and $\omegal$) and a
model for the radial mass distribution of the cluster. 
The X-ray or velocity data are usually
derived from the dense, central parts of 
the cluster ($r\simlt 500\h1$ kpc), whereas
the virial radius is generally in the range 1--$1.5\h1$ Mpc
for massive clusters. Consequently, an extrapolation
is entailed. Weak lensing data do yield mass as a function
of aperture that, in the case of MS1054, extend nearly
to the virial radius, but the mass in question is
a projected mass density, so that, again,
a model is required to obtain the virial
mass. 
These issues are not always well appreciated; in this section
we derive
$M_V$ in the context of a specific mass model
in order to make clear the assumptions involved.

We adopt the profile of Navarro, Frenk, \& White
(1997; NFW), which has been shown to be a good
fit to cluster-scale dark matter halos in N-body
distributions. The NFW profile has a density
distribution given by
\begin{equation}
\rho(r) = \frac{\rho_{crit}\, \delta_c}{x (1+x)^2}\,, 
\label{eq:rhonfw}
\end{equation}
where $x\equiv r/r_s$ and $r_s$ is a scale radius
for the halo. The central density is determined
by the parameter $\delta_c,$ which one expects
to be of order $10^4$ for massive clusters.
The mass interior to radius $r$ is then given by 
\begin{equation}
M(r) = 4\pi\rho_{crit}\,\delta_c\,r_s^3 \left[\ln(1+x)-\frac{x}{1+x}\right]\,.
\label{eq:mnfw}
\end{equation}  
Substituting equation~\ref{eq:mnfw} into the expression
for virial mass yields 
\begin{equation}
\frac{\ln(1+x_V) - \frac{x_V}{1+x_V}}{x_V^3} =
\frac{\Delta_V (1+z)^3\,\omegam}{3\,\delta_c}\,,
\label{eq:xv}
\end{equation}
where $x_V\equiv r_V/r_s.$ 

\subsection{Obtaining $M_V$ from X-ray temperature or velocity dispersion}
Now let us suppose that the observational
data consist either of a measurement of the
X-ray temperature $T_X$ of the intracluster gas,
or of the rms line-of-sight velocity dispersion 
of cluster galaxies, $\sigma_v.$ Applying
the equation of hydrostatic equilibrium to
the former, or the Jeans equation to the
latter, leads to the equation
\begin{equation}
\frac{G M(r)}{r^2}\,\rho_{X,g}\,
= \, -\frac{d}{dr}\left(\rho_{X,g} u^2\right)\,,
\label{eq:hydro}
\end{equation}
where $\rho_{X,g}$ is the density of the
X-ray emitting gas or the galaxy population
at radius $r,$ and
\begin{equation}
u^2  = 
 \left\{ \begin{array}{ll}
\frac{kT_X}{\mu\,m_p} \,, & {\rm X-ray\ gas\/} \,;\\
 & \\
\sigma_v^2\,,  & {\rm galaxy\ velocities}\,. \end{array} \right.
\end{equation}

We can solve equation~\ref{eq:hydro} if
we make the simplifying assumption that
the tracer (gas or galaxies) distribution
is roughly isothermal at the radii from which
most of the X-ray emission (galaxy velocities)
are derived, $r \approx r_s.$ I.e., we assume that
$d\ln\rho_{X,gas}/d\ln r \approx -2,$ $u^2 \approx$ constant.
This assumption is reasonable if the gas (galaxies)
approximately trace the dark matter potential,
because the NFW profile itself is nearly isothermal
at such radii. With the isothermal assumption
we obtain from equation~\ref{eq:hydro} the relation
\begin{equation}
\delta_c \approx 7\left(\frac{u}{H_0 r_s}\right)^2\,.
\label{eq:rsdc} 
\end{equation}

\vbox{%
\begin{center}
\leavevmode
\hbox{%
\epsfxsize=8.9cm
\epsffile{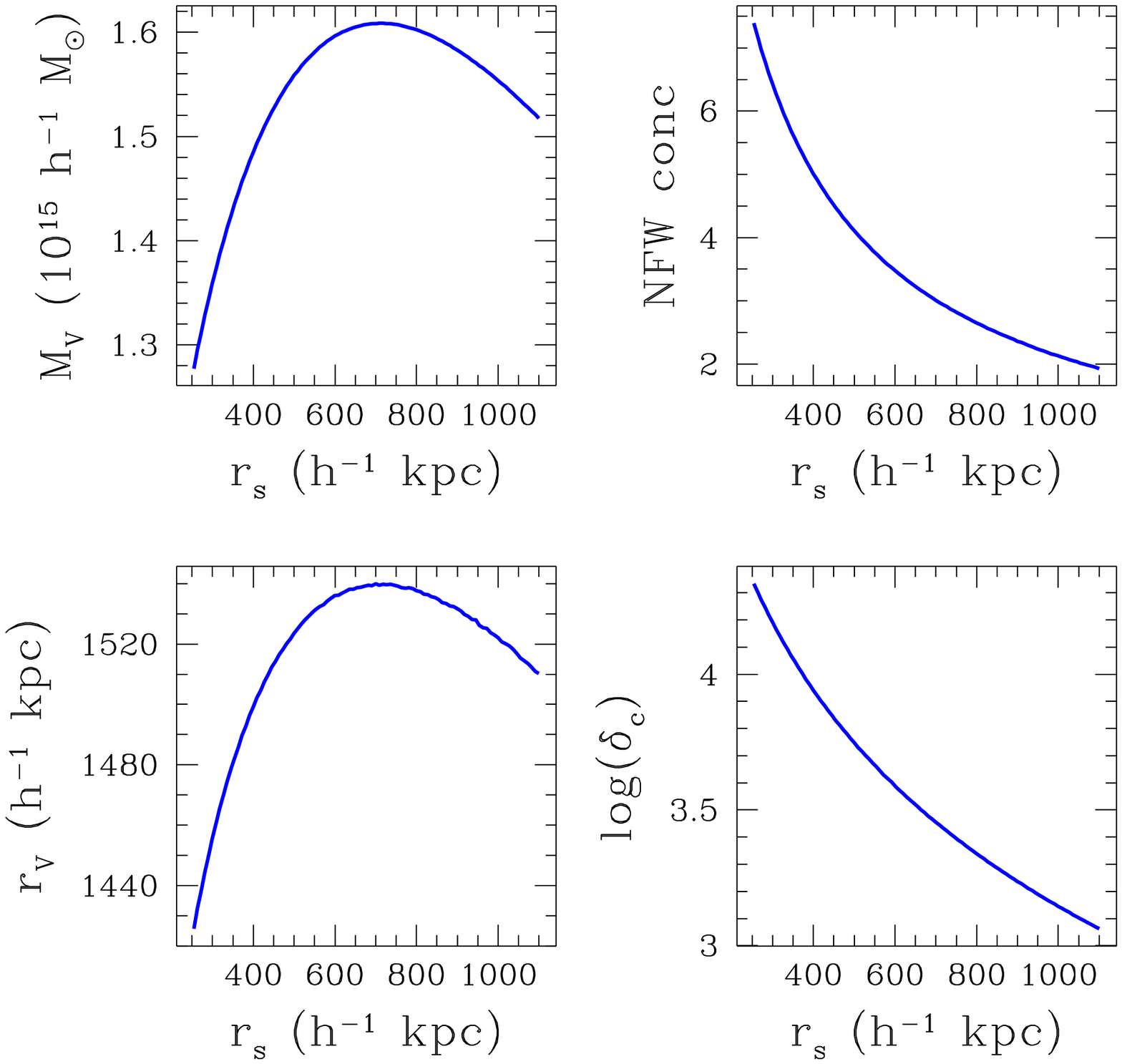}}
\begin{small}
\figcaption{%
The virial mass (upper left panel), 
virial radius (lower left panel),
NFW concentration parameter (upper right panel), and dimensionless
central density $\delta_c$ (lower right panel), for 
a cluster with an X-ray temperature of 12.3 keV,
plotted as a function of the adopted value of the NFW
scale radius $r_s.$ See text for details.
\label{fig:nfw_X}}
\end{small}
\end{center}}

Equation~\ref{eq:rsdc} shows that $\delta_c$ is determined from
observations (i.e., $T_X$ or $\sigma_v$) only if
the scale radius $r_s$ is known. In other words,
the two NFW parameters cannot be determined from
a single piece of information. Nevertheless, one might
hope that the virial mass itself is insensitive
to the particular values of $\delta_c$ and $r_s$ provided
they are related by equation~\ref{eq:rsdc}.
This indeed turns out to be the case, as we
show by adopting the following procedure.
First, we pick a value of $r_s$ in the range 250--1100\h1\ kpc.
From this we derive $\delta_c$ from equation~\ref{eq:rsdc}
and insert the result into equation~\ref{eq:xv} to obtain
$x_V.$ Substituting into equation~\ref{eq:mnfw}
then yields, after some algebra, the result
\begin{equation}
M_V =  \frac{21}{2}\,\frac{u^2}{G}\,r_s \left[\ln(1+x_V)-
\frac{x_V}{1+x_V}\right]\,.
\label{eq:mvir}
\end{equation}

Several aspects of equation~\ref{eq:mvir} 
merit further comment.
First, $x_V$ is 
a function of the observable $u$ and the adopted value of
$r_s.$ Thus, $M_V$ is not
quadratic in $u$ and linear in $r_s.$ Indeed,
the virial mass is virtually independent of
the adopted value of $r_s,$ as was argued above. This
is shown in the upper left panel of Figure~\ref{fig:nfw_X},
in which the $M_V$ is computed from an
X-ray temperature of $T_X=12.3$ keV, the value
obtained by Donahue \etal\ (1998) for MS1054 (cf.\ \S~\ref{sec:ms1054}).
(The figure assumes a flat, $\omegam=0.3$ cosmology
at $z=0.83,$ the redshift of MS1054.)
For $300 \simlt r_s \simlt 1000\h1$ kpc,
$M_V$ varies by less than \sm 10\%
from its mean value of $1.3\times 10^{15}\, \h1\msun$
for the adopted $T_X.$ The virial
radius $r_V$ changes by an even smaller percentage
over this range. In contrast, the inferred central
density $\delta_c$ is quite sensitive to 
the adopted $r_s.$ To constrain $r_s$ further,
the NFW concentration index $c$ is plotted
in the upper right hand panel. (This quantity
is not simply equal to $r_V/r_s,$ as it is
in Navarro \etal, because the virial
overdensity is not taken to be exactly 200.)
For massive clusters, the simulations of
Navarro \etal\ suggest that $c\simeq 3$--5.
Thus, the plausible values
of $r_s$ are in the range $300 \simlt r_s \simlt 700\h1$ kpc.
As we have seen, $M_V$ varies little for $r_s$ in this
range. Consequently, we can be assured that
our procedure introduces less than about 10\% error
in the virial mass.

Second, the virial mass as determined from
equation~\ref{eq:mvir} differs from what
would be obtained had we assumed (as is
usually done) a purely isothermal
structure for the cluster. Specifically,
the virial mass obtained from the NFW
model is 10--20\% {\em larger\/} than
an isothermal mass estimate, with the
precise factor depending on the typical
concentration index. In a sense it is
reassuring that the difference
is relatively small; on the other
hand, as we shall see below, even small
mass differences can be crucial
in terms of inferences about nongaussianity.
Constraining the mass profiles of clusters
is thus an important ongoing task
for cluster physics.

Last, it is important to bear in mind that
$M_V$ as derived from equation~\ref{eq:mvir}
is dependent on the adopted cosmology and
redshift, via equation~\ref{eq:xv}, from
which $x_V$ is derived. Thus, {\em the virial
mass of a cluster, derived from temperature
or velocity observations, cannot be specified
independently of adopted cosmology.} This
property of the virial mass, as defined by
the PS formalism, will be important in the
analysis presented below.

%
%

%
%

%
%
%
%
%

%
%
\subsection{Obtaining $M_V$ from weak lensing data}

Weak lensing data constitute the third and, in principle,
most rigorous means of deriving cluster masses, as
there is no need to assume hydrostatic or dynamical equilibrium.
Deriving a virial mass given 
aperture mass measurements nonetheless requires
care, as we now discuss.

\def\kbar{{\overline \kappa}}
The quantity measured directly by a weak
lensing shear analysis (e.g., Luppino \& Kaiser 1996)
is the mean dimensionless surface density $\kbar$
within an angular radius $\theta$ of the cluster center.
The projected mass within $\theta$ is then given by
\begin{equation}
M(\theta) = \kbar \frac{c^2}{4\pi G} \frac{D_S D_L}{D_{LS}} \pi\theta^2\,,
\label{eq:mtheta}
\end{equation}
where $D_L,$ $D_S,$ and $D_{LS}$ are the lens, source, and lens-source
angular diameter distances respectively. These
angular diameter distances are all cosmology-dependent;
moreover, the typical redshift of the lensed sources must
be assumed. These factors all influence the derived
virial mass.
Evaluating numerical factors in equation~\ref{eq:mtheta}
yields
\begin{equation}
M(\theta) = 2.13\times 10^{15} \frac{\kbar}{0.1} \left(\frac{\theta}
{4'}\right)^2 \frac{y_L y_S}{(1+z_L) y_{LS}} \h1\msun\,, 
\label{eq:mtheta1}
\end{equation}
where the $y$'s are the dimensionless comoving angular diameter
distances as defined by Peebles (1993).

To make the connection with the NFW halo parameters,
we must equate the observationally derived aperture mass
above with the projected mass calculated from
the NFW formulae; we show elsewhere (Willick \& Padmanabhan 1999)
that this is given by
\begin{equation}
M(\theta;\delta_c,r_s) = 4\pi\rho_{crit}\delta_c r_s^3 \beta(D_L\theta/r_s)\,,
\label{eq:mthetanfw}
\end{equation}
where 
\begin{equation}
\beta(x) = \int_0^x y\,dy \int_0^\infty (y^2+w^2)^{-1/2} (1+y^2+w^2)^{-1}\,dw
\end{equation}
is the projected (dimensionless) mass density for the NFW
profile. (Explicit expressions for $\beta$ 
are given by Willick \& Padmanabhan 1999.) Equating the lensing-inferred
aperture mass (equation~\ref{eq:mtheta1}) with the NFW model
aperture mass (equation~\ref{eq:mthetanfw})
yields the following expression for $\delta_c$ as
a function of $r_s,$ $\kbar,$ and $\theta:$
\begin{equation}
\delta_c= 4886 \frac{\kbar}{0.1} \left(\frac{\theta}
{4'}\right)^2 \beta^{-1} \left(\frac{r_s}{500\ {\rm kpc}}\right)^{-3}
\frac{y_L y_S}{(1+z_L)y_{LS}}\,.
\label{eq:dclens}
\end{equation}

\vbox{%
\begin{center}
\leavevmode
\hbox{%
\epsfxsize=8.9cm
\epsffile{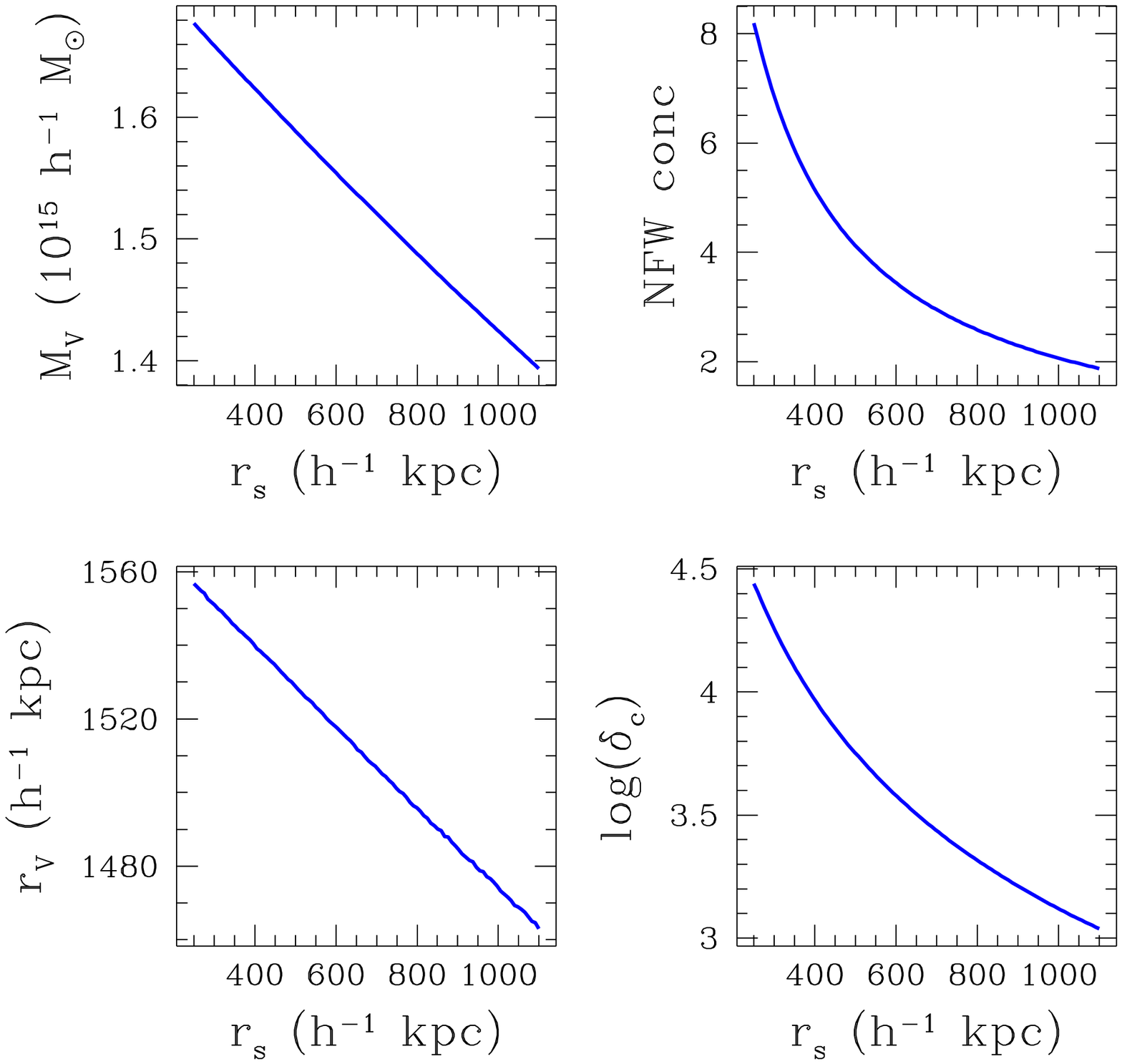}}
\begin{small}
\figcaption{%
Same as the previous figure, except that now
the virial mass and related quantities are
derived from lensing data. A mean convergence
of $\kbar=0.08$ within $\theta=4',$ the values
for MS1054, have been used in the calculation.
\label{fig:nfw_L}}
\end{small}
\end{center}}
Equation~\ref{eq:dclens} is the analogue of equation~\ref{eq:rsdc}
for the case of a lensing measurement; it gives $\delta_c$ from
the data for any adopted value of $r_s.$ This suggests that we
proceed as above: pick a value of $r_s$ within a sensible
range; calculate the corresponding $\delta_c$ from equation~\ref{eq:dclens},
and thus the virial radius in units of $r_s,$ $x_V,$ from equation~\ref{eq:xv};
obtain the virial mass $M_V=4\pi r_V^3 \Delta_V\overline\rho/3.$
We expect that, as found in the temperature/velocity dispersion
case, the virial mass will be insensitive to the choice of $r_s.$
Figure~\ref{fig:nfw_L}, in which virial mass, radius,
NFW concentration index, and central density are plotted
for the lensing parameters of MS1054 (see next section),
shows that this is once again the case. 

\section{The Virial Mass of MS1054}
\label{sec:ms1054}
We now apply the methods of the previous section
to estimate the virial mass of MS1054 as a function
of $\omegam.$ We assume a flat universe, $\omegam+\omegal=1,$
so that the quantity $\Delta_V(z)$ discussed in
the previous section is fully determined by $\omegam.$

MS1054 lies at a redshift of $z=0.833$ (Tran \etal\ 1999, T99),
making it one of the most distant confirmed rich clusters.
Its mass can be estimated via each of the three principal
methods of the previous section.
T99 found it to have a velocity dispersion of $1170\pm 150\ \kms$
based on 24 galaxies. 
An earlier determination of the velocity
dispersion by Donahue \etal\ (1998; D98) found 
$\sigma_v=1360\ \kms$ based on 12 galaxies.
D98 also obtained ASCA data for MS1054, from which they deduced
a temperature of 
$T_X=12.3^{+3.1}_{-2.2}$ keV for the intracluster gas,
equivalent
to a velocity dispersion $kT_X/\mu m_p=(1413\ \kms)^2$
for $\mu=0.59,$ the value adopted by Borgani \etal\ (1999).
Luppino \& Kaiser (1997, LK97) obtained
$V$ and $I$ band images of MS1054 at the 2.2 m telescope
on Mauna Kea in good ($\sim 1''$) seeing, and used these
data to estimate $\kbar(\theta).$ Their mass estimates (for
which they assumed $\omegam=1$) indicate a $\kbar=0.1$ for
$\theta=4$ arcmin. However, a more conservative estimate,
obtained from their Figure 7, is $\kbar=0.08$ for $\theta=4',$
and we use this more conservative estimate here.

Using these data we can determine the virial mass of
MS1054 for an adopted cosmology. Because of the $\sim 10\%$
variations in the deduced mass with the adopted value of $r_s,$
we must impose an additional constraint. The most reasonable
one to adopt is to require the NFW concentration parameter $c$
to take on a particular value. Here we choose $c=4.$ For the
lensing mass, an additional assumption is required, namely,
the mean redshift, $z_S,$ of the lensed galaxy population. Here we
choose $z_S=1.5,$ which LK97 considered the most suitable
value, and which yields mass estimates midway between the
minimum ($z_S\approx 1$) and maximum ($z_S\approx 3$) allowed
values (LK97). 

Figure~\ref{fig:ms1054_mass} shows the virial masses obtained
via each of the three methods as a function of $\omegam.$
The cosmological dependence of the virial mass is clear. 
We have also calculated, for use in the next section,
a weighted average mass and the \onesigma\ errors on 
this average. Mass errors were estimated
by considering the principal sources of
error for each method: velocity dispersion uncertainty for the dynamical
mass, temperature uncertainty for the X-ray mass,
and source redshift uncertainty for the weak lensing
mass. The corresponding mass errors were estimated
to be 42\%, 35\%, and 40\% respectively. The individual
masses were weighted inversely with the squares
of the fractional mass errors to obtain the average mass,
which is shown as a heavy solid curve in the plot;
the shaded region around the average shows its $\pm 1\,\sigma$
uncertainty, which is $\sim 22\%.$

As the figure shows, while the X-ray and
lensing masses agree to within \sm 10\% for $\omegam\simlt 0.5,$
the dynamical mass is 40--50\% smaller than the other two.
This difference is within the observational errors, and
thus is not indicative of systematic differences. Nonetheless, the
sensitivity of the PS-predicted cluster abundance to mass makes this
difference potentially significant. We discuss this important
issue further in \S~7.2

\vbox{%
\begin{center}
\leavevmode
\hbox{%
\epsfxsize=8.9cm
\epsffile{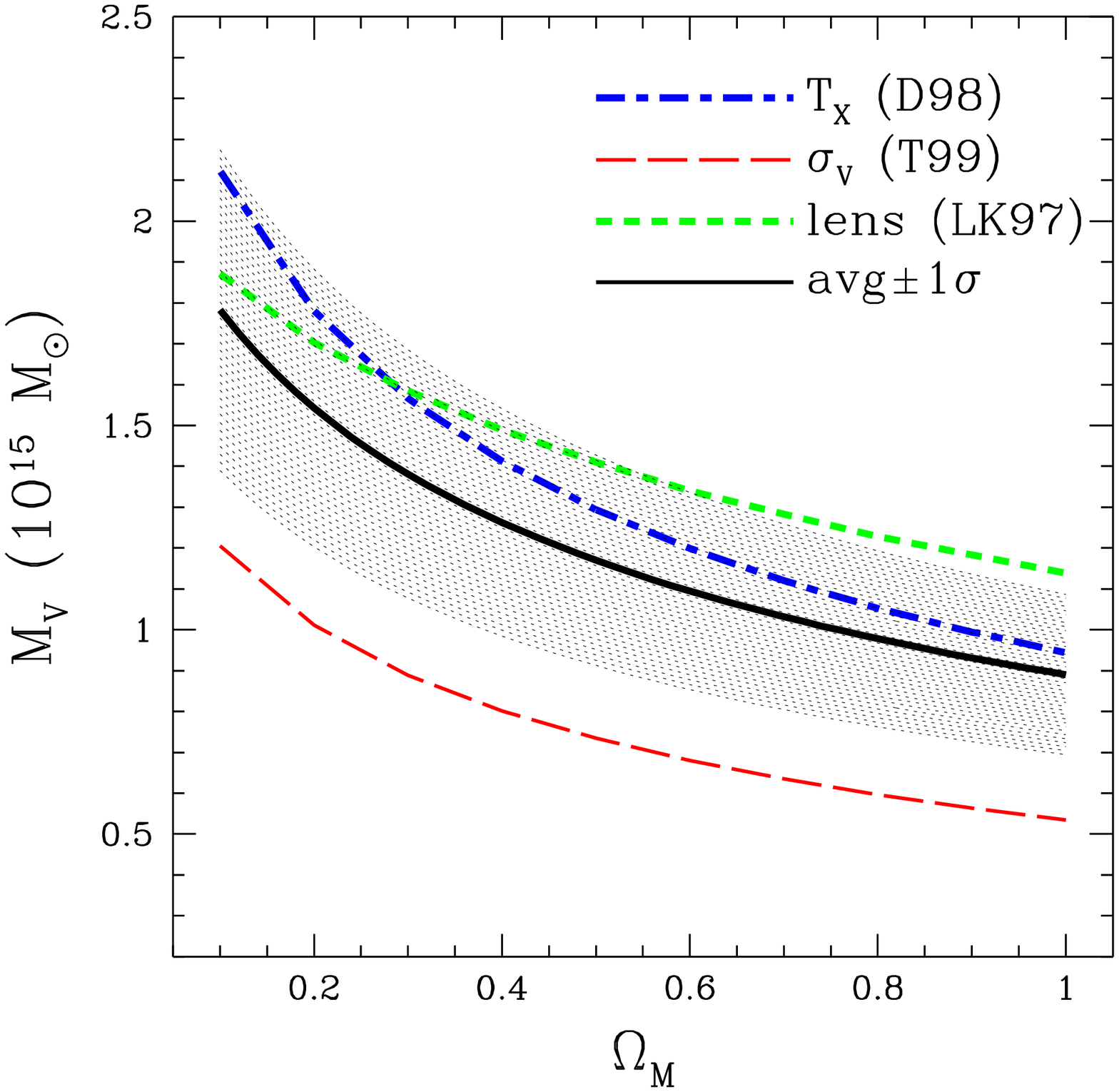}}
\begin{small}
\figcaption{%
The virial mass of MS1054 computed from X-ray, galaxy velocity,
and lensing data, plotted as a function of $\omegam.$
The weighted average is plotted as a solid black
line, with the light shading showing the $\pm 1\,\sigma$
errors.
\label{fig:ms1054_mass}}
\end{small}
\end{center}}

\section{Method of Analysis and Basic Results}
\label{sec:results}
\def\sig8{\sigma_8}
In order to apply the PS formalism to high-redshift
clusters, one must first normalize the mass fluctuations
on cluster scales by requring the PS prediction
to match observations
at {\em low\/} redshift. 
The result is generally expressed as a relationship
between $\sig8,$ the value of $\sigma_M$ within
a top-hat sphere of radius $8\h1$ Mpc, and $\omegam.$
Recent results obtained from X-ray temperature
and luminosity data include those of Borgani \etal\ (1999):
\begin{equation}
\sig8 = (0.58\pm 0.06) \times \omegam^{-0.47+0.16\omegam}\,;
\label{eq:B99}
\end{equation}
Wang \& Steinhardt (1998):
\begin{equation}
\sig8 = (0.50\pm 0.10) \times \omegam^{-0.43-0.33\omegam}\,;
\label{eq:BWS98}
\end{equation}
and Pen (1998):
\begin{equation}
\sig8 = (0.53 \pm 0.05) \times \omegam^{-0.53}\,.
\label{eq:P98}
\end{equation}
(In each case, results for a flat universe, $\omegam+\omegal=1,$
are given; in the case of Wang \& Steinhardt (1998) nominal
values of other cosmological parameters have been adopted.
See the original papers for further details.)
Similar expressions 
have been given by Eke, Cole, \& Frenk (1996), Girardi \etal\ (1998),
and Suto \etal\ (1999).

\vbox{%
\begin{center}
\leavevmode
\hbox{%
\epsfxsize=8.9cm
\epsffile{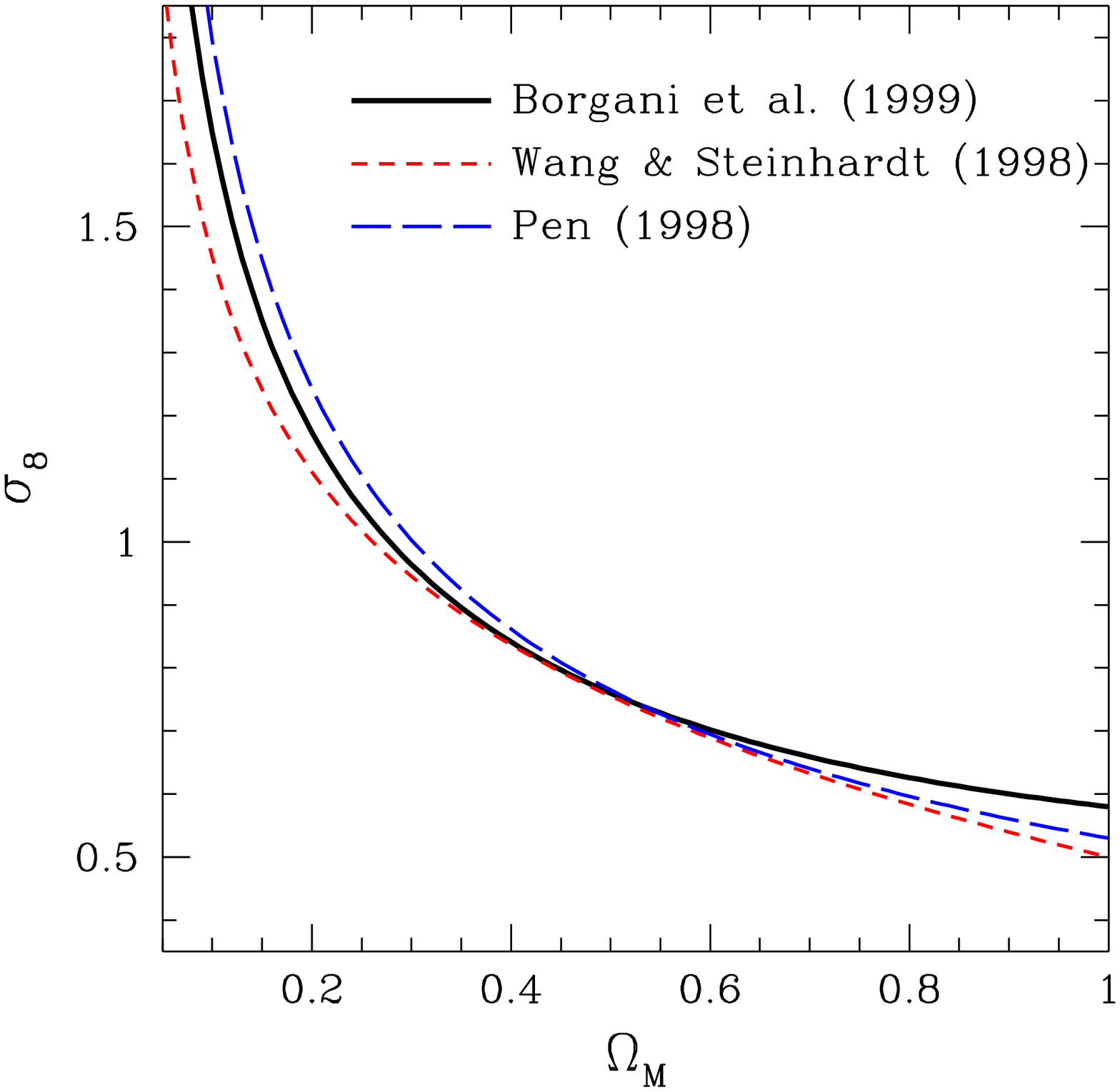}}
\begin{small}
\figcaption{%
Three recently published $\sig8$-$\omegam$ relations
derived from low-redshift cluster X-ray temperature and luminosity
data. 
\label{fig:sig8omegam}}
\end{small}
\end{center}}

Figure~\ref{fig:sig8omegam} shows $\sig8$ as
a function of $\omegam$ as given by equations~\ref{eq:B99},
~\ref{eq:BWS98}, and~\ref{eq:P98}. 
The agreement
is within the reported errors, and is especially good
for $0.4\simlt \omegam \simlt 0.7.$ 
However, because of the exponential sensitivity
of the PS abundance predictions to $\sigma_M,$
the small differences in figure~\ref{fig:sig8omegam}
have a nonnegligible effect on the inferred degree
of nongaussianity. In what follows, we adopt the
Borgani \etal\ (1999) calibration, which
is intermediate between the other two for
$\omegam$ in the observationally preferred
range (\sm 0.2--0.5). Furthermore, it is
based on the most extensive and
recent compilation of X-ray data, on which
the conclusions of the present paper heavily
depend.

\subsection{Effect of nongaussianity on the $\sig8$-$\omegam$ relation}

The above low-redshift $\sig8$-$\omegam$ relations were, of
course, derived under the assumption of Gaussian primordial
density perturbations. If we relax this assumption---e.g.,
assume the fluctuations are described by the $\psi_\lambda$
distribution for finite $\lambda$---we expect the
$\sig8$-$\omegam$ relation to change as well. 
To obtain the relation for finite $\lambda$
we first note that the 
relations essentially represent the requirement
that the $z=0$ PS abundance prediction, for
any $\omegam,$ match
the observed abundances 
at a characteristic rich cluster mass. 
Specifically, the Borgani \etal\ (1999)
calibration yields $M n(M) = 4.2 \times 10^{-6}\ \h1\ {\rm Mpc}^{-3}$
for $M = 5.95 \times 10^{14}\,\h1\msun$ (i.e., for $m=1$). 
By requiring that this condition hold for finite
$\lambda$ as well, we obtain modified $\sig8$-$\omegam$
relations for any desired degree of nongaussianity.

\vbox{%
\begin{center}
\leavevmode
\hbox{%
\epsfxsize=8.9cm
\epsffile{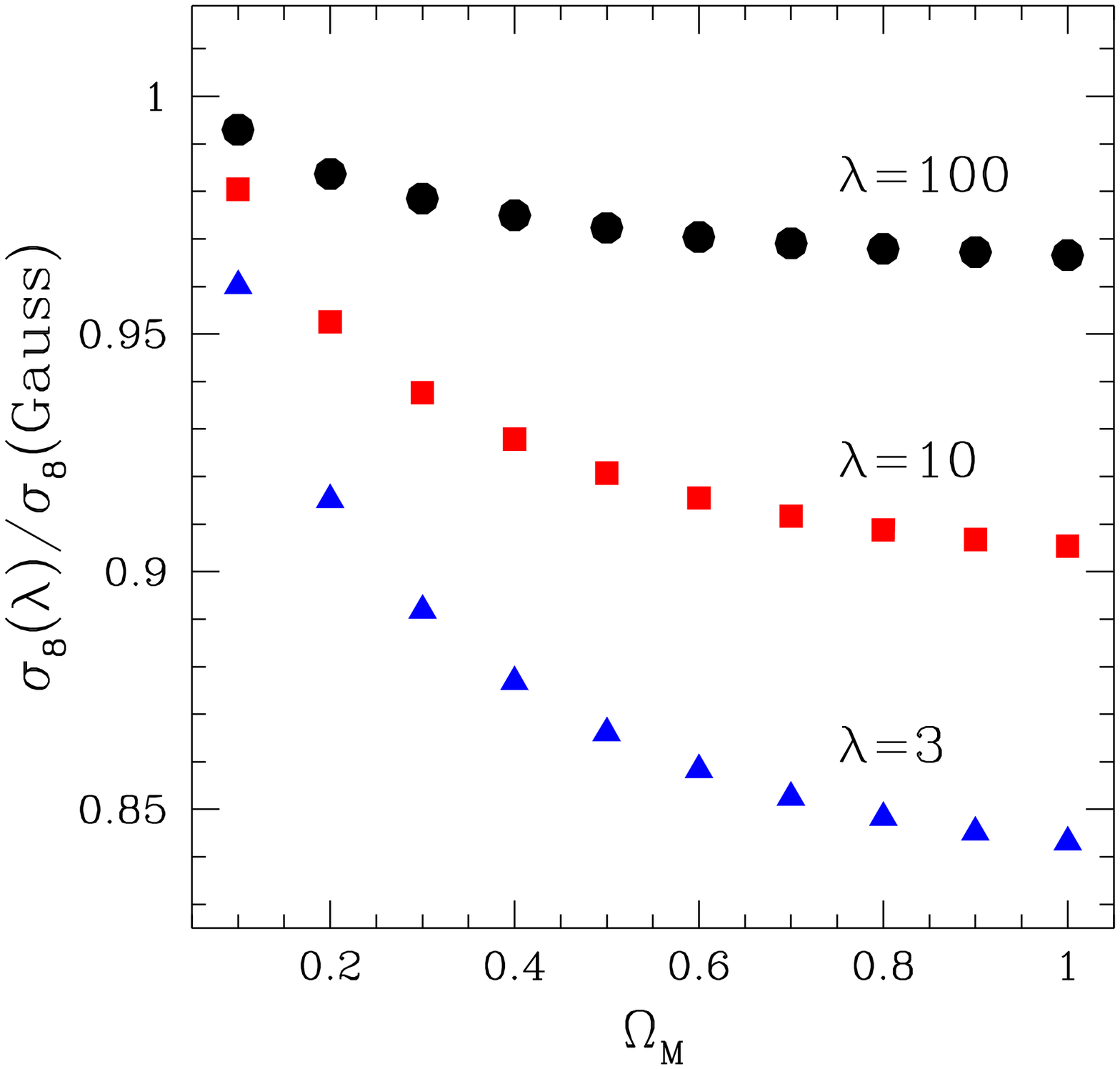}}
\begin{small}
\figcaption{%
The ratio of $\sig8$ for the $\psi_\lambda$ distribution
to the Borgani \etal\ value of $\sig8,$ plotted as
a function of $\omegam.$ The values of $\sig8(\lambda)$
were obtained by requiring the generalized PS abundance
of $M=6\times 10^{14}\,\h1\ \msun$ clusters to match that
obtained for the Gaussian case, at each value of $\omegam.$
\label{fig:s8sl}}
\end{small}
\end{center}}

Figure~\ref{fig:s8sl} shows the effect on the $\sig8$-$\omegam$
relation of using the generalized PS abundance formula, 
equation~\ref{eq:psgen}, with the $\psi_\lambda$ distribution
for $\lambda=3,$ $10,$ and $100.$ It can be seen that
the changes relative to the Gaussian case are rather small,
especially for $\omegam\leq 0.2.$ This is because low-redshift
clusters of moderate mass do not represent very high
peaks in the initial density field, for low $\omegam.$
For significant nongaussianity $\lambda\simlt 3$ and
$\omegam\simgt 0.4,$ the fluctuation normalization
is more subtantially modified.

\subsection{Calculating the expected number of
MS1043-like clusters}

Once we have modified the normalization of the
density fluctuations $\sigma_M$ for a given
value of $\lambda$ as described above, we may
calculate the predicted comoving number
density of clusters above a given mass
threshold using equation~\ref{eq:numden_gen}
with $\psi=\psi_\lambda.$
When Gaussian fluctuations are assumed, we
use equation~\ref{eq:numden}. To determine
the number of clusters, of mass and redshift as large
or larger than that of MS1054, 
expected in
the EMSS sample, we integrate this number
density over redshift, multiplying by the
appriate comoving volume element:
\begin{equation}
{\cal N}_{exp} = \omega_{1054} \int_{z_1}^{z_{max}} 
N(\ge\!m_{1054},z) \frac{dV}{dz}\,dz\,,
\label{eq:nexp}
\end{equation}
where:
\begin{enumerate}
\item $m_{1054}$ is the virial mass, in dimensionless units
(cf.\ \S~2.2), of MS1054 as determined by the methods
of \S~\ref{sec:clmass};
\item $\omega_{1054}$ is the solid angle covered
by the Einstein satellite to a limiting X-ray flux
fainter than the observed flux of MS1054. From Tables 1 and 2
of Henrey \etal\ (1992) we find $\omega_{1054}=0.041$ sterradian;
\item $z_1=0.833$ is the redshift of MS1054 (T99);
\item $z_{max}$ is the maximum redshift out to which
a cluster whose X-ray luminosity is equal to that of
MS1054 could have been detected by the EMSS. In practice,
$N(\ge\!m,z)$ is dropping so rapidly with increasing
redshift that the result is insensitive to $z_{max}$
for $z_{max}\simgt 1,$ and we conservatively set
$z_{max}=1.3$ in the calculations to follow;
\item $dV/dz$ is the cosmology-dependent comoving
volume per unit redshift (e.g., equation~13.61 of
Peebles 1993).
\end{enumerate}

\vbox{%
\begin{center}
\leavevmode
\hbox{%
\epsfxsize=8.9cm
\epsffile{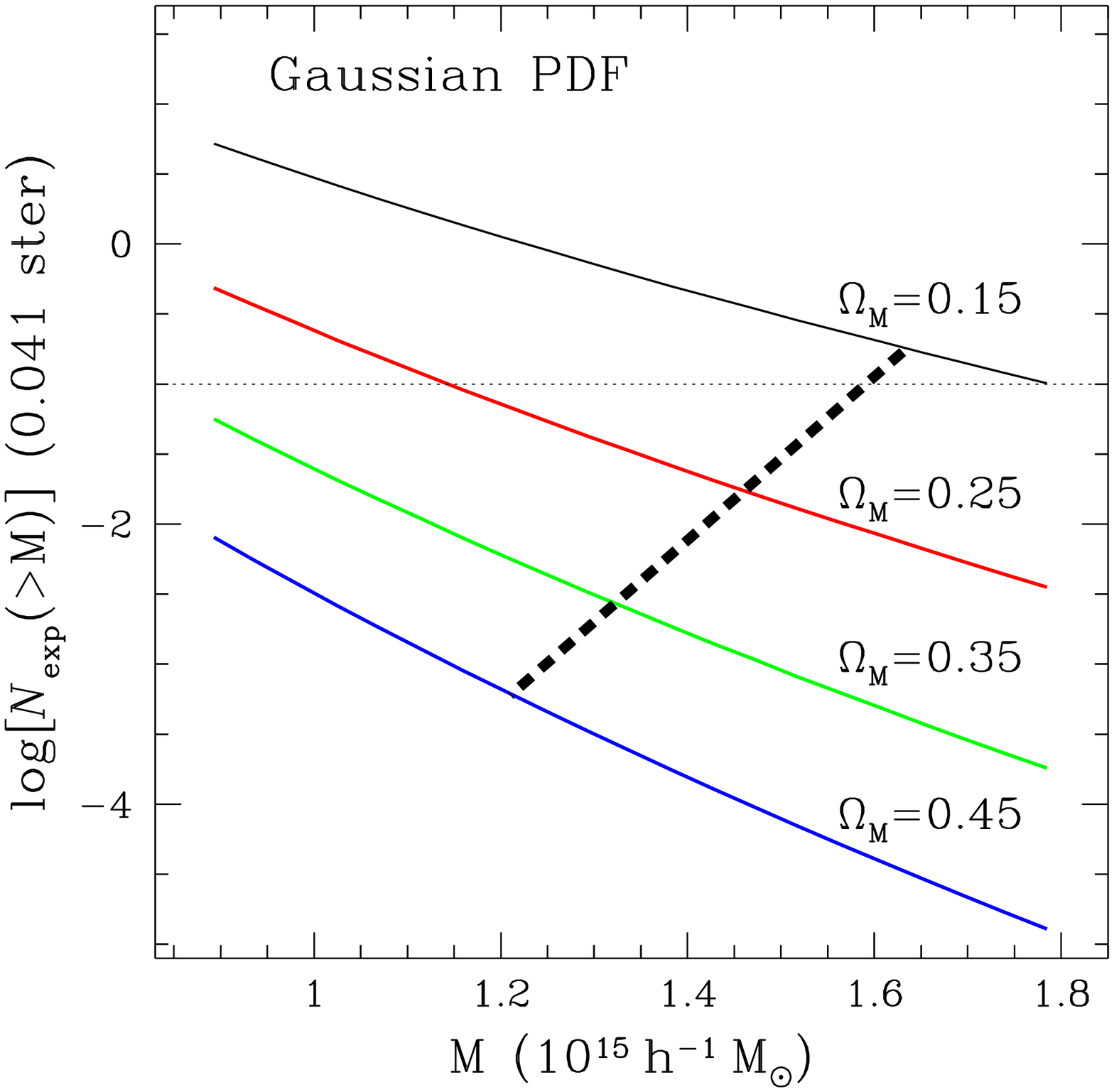}}
\begin{small}
\figcaption{%
Expected number of MS1054-like clusters in the EMSS survey,
plotted as a function of mass, under the assumption
of Gaussian initial density fluctuations. The $\sigma_8$-$\omegam$
relation is that of Borgani \etal\ (1999), and a flat
universe is assumed. The different curves are for four
different values of $\omegam.$ The heavy dashed line sloping
down from right to left indicates the variation of the virial mass
of MS1054 with $\omegam$ (see Figure~\ref{fig:ms1054_mass}).
The horizontal dotted line indicates a one in ten chance
that a cluster as massive as MS1054 would be present in the
EMSS survey. Note that only this occurs only for $\omegam=0.15.$
\label{fig:nexp_gauss}}
\end{small}
\end{center}}

Figures~\ref{fig:nexp_gauss} and~\ref{fig:nexp_lambda} show the
results of calculating ${\cal N}_{exp}$
for a Gaussian PDF and a $\lambda=1.5$ $\psi_\lambda$-PDF
respectively. Each figure plots ${\cal N}_{exp}$ for
four representative values of $\omegam.$ The heavy dashed
line sloping down and to the left indicates how the
virial mass of MS1054 changes with $\omegam$ (cf.\ \S~\ref{sec:ms1054}).
Where that line intersects the curve for a given $\omegam$
yields the predicted number of MS1054-like clusters in the EMSS.
As Figure~\ref{fig:nexp_gauss}
shows, this number is quite small for all values
of $\omegam\ge 0.25$ in the Gaussian case.
A cluster at $z=0.83,$ and as massive as the
lensing, X-ray, and velocity data for MS1054 indicate
that it is, is unlikely to have been found if the
fluctuations are Gaussian, unless $\omegam\simlt 0.2.$

The situation is significantly changed for a
$\psi_\lambda$-PDF with $\lambda=1.5.$ Figure~\ref{fig:nexp_lambda}
shows that an MS1054-like cluster now has a better than 10\%
chance of being found for $\omegam\simlt 0.4.$ Thus, nongaussian
fluctuations allow a much larger value of $\omegam$
to be consistent with the MS1054 data.

\vbox{%
\begin{center}
\leavevmode
\hbox{%
\epsfxsize=8.9cm
\epsffile{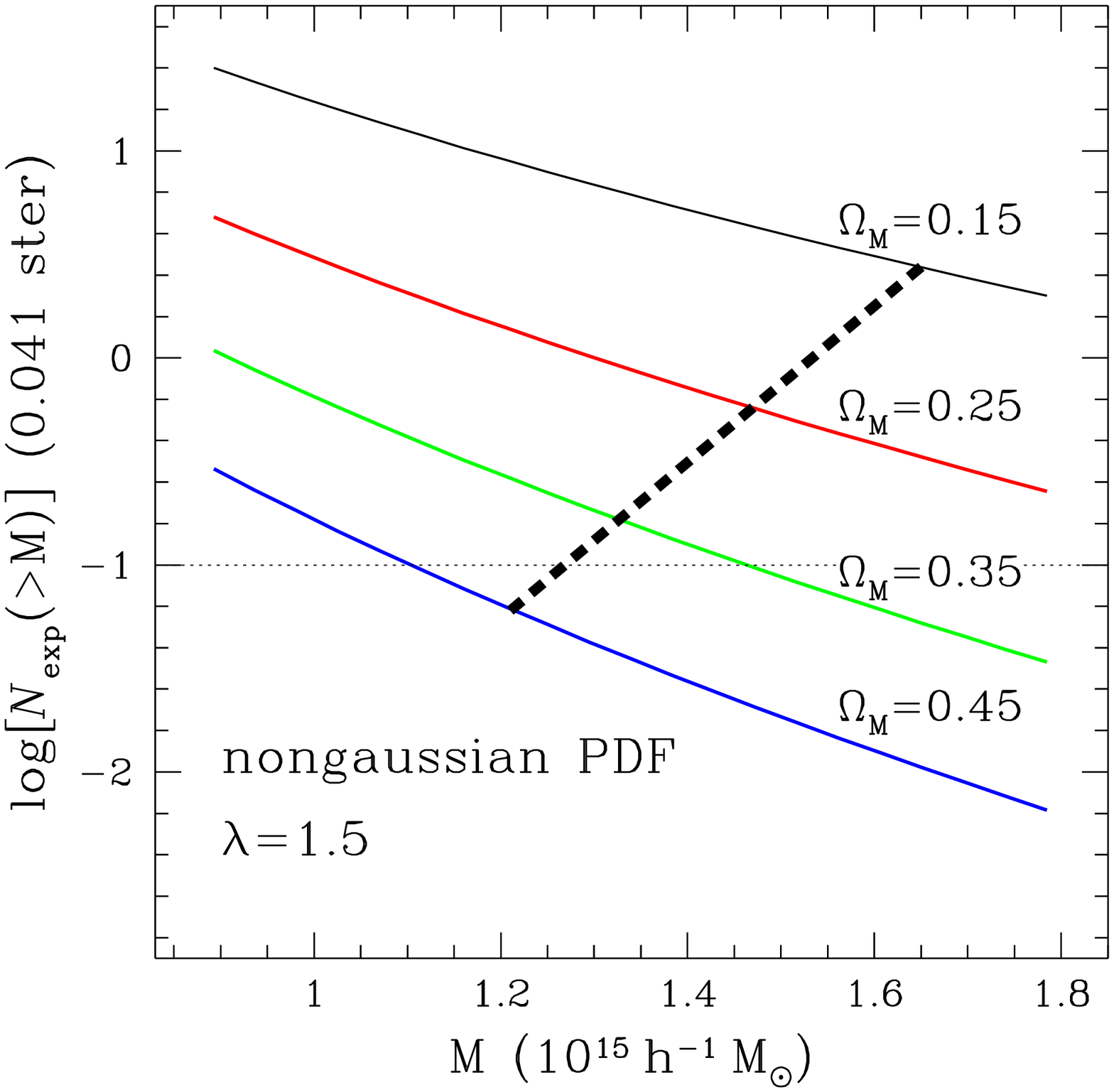}}
\begin{small}
\figcaption{%
Same as the previous figure, except that now
the expected number of clusters is calculated
using a PDF described by the $\psi_\lambda$
distribution, with $\lambda=1.5.$
\label{fig:nexp_lambda}}
\end{small}
\end{center}}

Because we are dealing with a sample of one, we cannot
hope to estimate a value of $\lambda$ per se. It is,
however, reasonable to derive an upper limit on $\lambda$
(i.e., a lower limit on the required degree of nongaussianity)
as a function of $\omegam.$ To do so, we calculate, for
each $\omegam,$ the value of $\lambda$ required to
yield ${\cal N}_{exp}=0.1,$ i.e., a one in ten chance
that an MS1054-like cluster would be found in the EMSS
for that value of $\omegam.$ The resultant value of
$\lambda$ may
be thought of as a 90\% confidence level upper
limit on $\lambda.$ The corresponding value of
the $T$-statistic (see Figure~\ref{fig:Tpsi})
would then be a 90\% confidence level lower limit.

Figure~\ref{fig:lambda_T} shows the results of carrying
out this calculation, and thus summarizes the main
results of this paper. The upper limit on $\lambda$
decreases from $\infty$ (Gaussian fluctuations) for
$\omegam=0.17,$ to \sm 100 for $\omegam=0.20,$
to much smaller values for $\lambda\simgt 0.25.$
As the upper limit on $\lambda$ decreases with
increasing $\omegam,$ signifying increasing
nongaussianity, the lower limit on $T$ increases,
from 1 (Gaussian fluctuations) at $\omegam=0.17,$
to \sm 2 at $\omegam=0.25,$ to $\simgt 6$ for
$\omegam\ge 0.4.$ The required amount of
nongaussianity rapidly increases with increasing
$\omegam.$ This is ultimately a reflection of
of the later ``freeze-out'' time for fluctuation
growth in higher density universes.

\vbox{%
\begin{center}
\leavevmode
\hbox{%
\epsfxsize=8.9cm
\epsffile{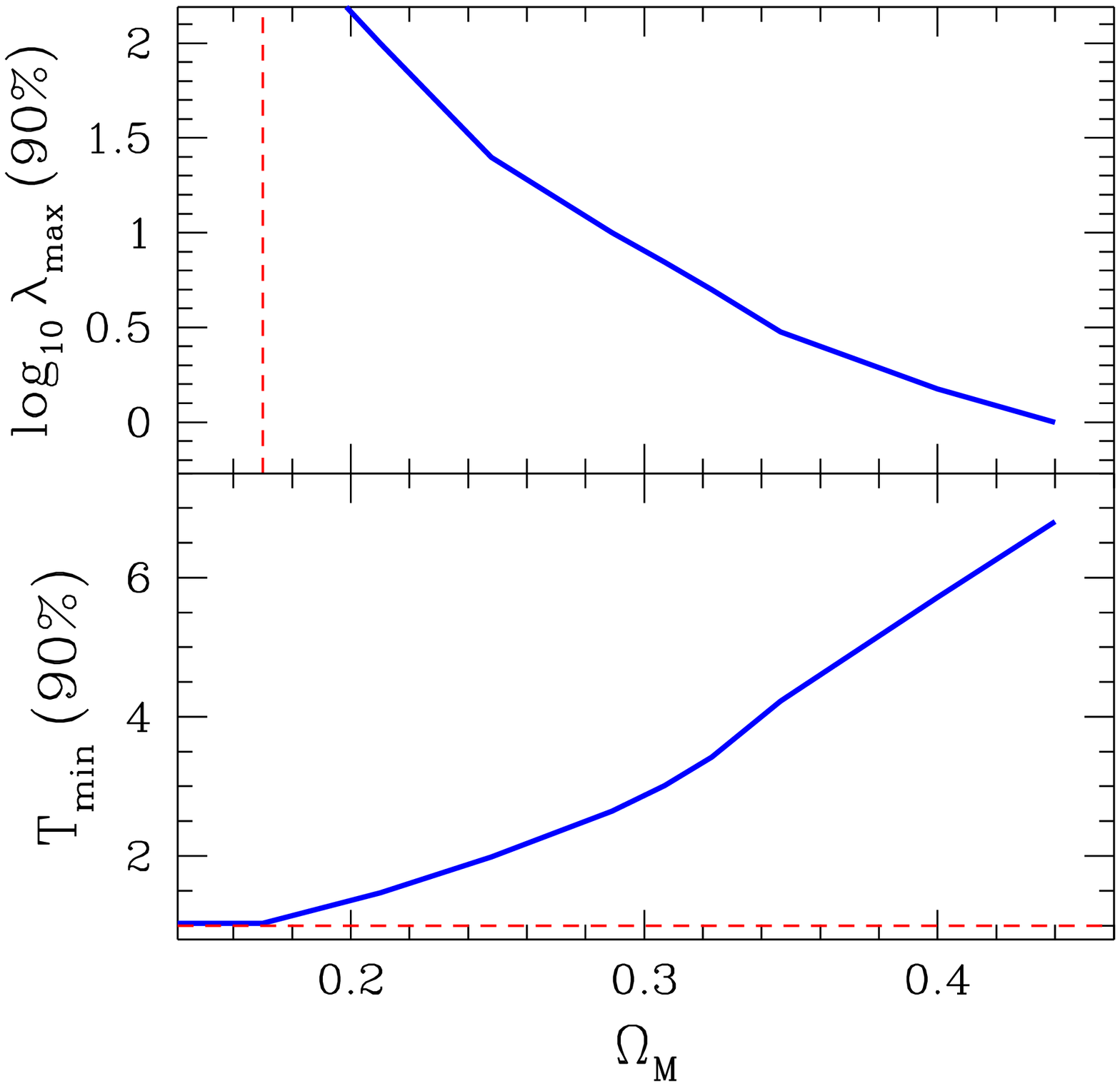}}
\begin{small}
\figcaption{%
The upper panel shows the 90\% confidence level
upper bound on $\lambda$ as a function of $\omegam.$
The lower panel shows the corresponding {\em lower\/}
bound on $T.$ These 90\% bounds correspond to the
amount of nongaussianity required for the expected
number of MS1054-like clusters in the EMSS to be $\ge 0.1.$ 
See main text for further details.
\label{fig:lambda_T}}
\end{small}
\end{center}}

\section{Discussion}
\label{sec:disc}
We have shown that the X-ray selected cluster
MS1054--03, at $z=0.83,$ adds to a small but growing body
of evidence that the primordial density fluctuation field
may be nongaussian. Specifically, the high mass of
MS1054, $\simgt 10^{15}\,\h1\msun,$ indicates that rare
fluctuations, $\delta/\sigma_M\geq 3,$ are more probable at
early times than they would be in the Gaussian case,
if $\omegam\simgt 0.2.$ In this concluding section, we
further discuss several aspects of this result.

\subsection{Comparison with previous work}

\subsubsection{Previous analyses based on MS1054}

Two other recent papers have considered the cosmological
implications of MS1054: D98 (cf.\ \S~\ref{sec:ms1054}) and
Bahcall \& Fan (1998; BF98).  Both papers reached the conclusion
that, by virtue of its high mass, MS1054 by itself rules
out an Einstein-de Sitter universe at a high confidence
level. D98 stated that MS1054 is consistent with a flat
$\omegam=0.3$ cosmology, while BF98
found that MS1054 is most consistent with $\omegam=0.2.$
Our results are in rough accord with theirs, although
for Gaussian fluctuations we find $\omegam\le 0.17,$ somewhat
smaller than BF98 and markedly lower than D98. The latter
difference may result from D98's conservative
estimate of the virial mass of MS1054, $\sim 7\times 10^{14}\,\h1\msun,$
which assumed $\omegam=1$ and an isothermal mass model. In any case,
comparison with D98 and BF98 is imprecise
because they did not consider nongaussianity as an additional
degree of freedom in the abundance analysis; their constraints
on $\omegam$ are valid only if the initial density field is Gaussian.

\subsubsection{Other cluster analyses allowing nongaussianity}

A more direct comparison of our results may be made with the
recent papers by Robinson, Gawiser,
\& Silk (1998; RGS98) and Koyama, Soda, and Taruya (1999; KST99).
The main points of those papers were summarized in \S~\ref{sec:intro};
here we discuss them further in light of our findings.

RGS98 and KST99 
considered not only the cluster abundance,
but also the cluster correlation length, using observational
constraints on the latter obtained
from the APM survey (Croft \etal\ 1997), and compared with the
predictions of the generalized PS formalism. While increasingly nongaussian
PDFs {\em increase\/}
the predicted cluster abundance (for given $\omegam$ and $\sigma_8$),
they {\em decrease\/} the predicted correlation length. This helps
break the degeneracy between $\omegam$ and nongaussianity (e.g., $T$)
present in any analysis which, like the present one, considers
only abundance data. In particular, while our analysis
allows Gaussian fluctuations for sufficiently
low $\omegam,$ RGS98 and KST99 found that large $T$
is in fact required for low ($\simlt 0.2$) $\omegam$
in order to accommodate
both abundance and correlation length data.

The analyses of RGS98 and KST99 differed in that only the
latter used intermediate-redshift cluster abundance
as an observational constraint (each considers $z\simlt 0.1$
abundances and correlations).  This makes KST99 the more
powerful probe of parameter space. While 
RGS98 found an Einstein-de Sitter universe with a Gaussian PDF
to be a good fit to the data, KST99 ruled out $\omegam\ge 0.5.$
They found that the required degree
of nongaussianity is a minimum for $\omegam\simeq 0.3,$ at which
$T=3.8\pm 1.0.$ From Figure~\ref{fig:lambda_T} we see that for
$\omegam\simeq 0.3$ our study indicates $T \ge  3,$
consistent with the KST99 result.
For a flat $\omegam=0.3$ universe with $\Gamma=0.20,$ RGS98 obtained
$T=4.0^{+3.6}_{-2.0},$ consistent with both KST99 and this paper.

\subsection{On the form of nongaussian PDF}

KST99 assumed $P(\delta|M) \propto \chi^2_m(\delta/\sigma_M),$ 
whereas we have taken $P(\delta|M) \propto 
\psi_\lambda(\delta/\sigma_M).$ The two PDFs differ more in spirit
than in practice. The $\chi^2_m$ distribution has its roots in
a physical model, in which the CDM is produced at the end
of an inflationary epoch from $m$ scalar fields which
couple quadratically to a Gaussian inflaton field. The $\psi_\lambda$
distribution, on the other hand, has no particular physical motivation,
being simply a mathematical transformation of the familiar
Poisson distribution (\S~\ref{sec:nongauss}). As we showed
in \S~\ref{sec:nongauss}, 
the two distributions are remarkably
similar in terms of their $T$-statistic, provided one makes
the identification $m=9\lambda.$ In fact, $\chi^2_{m=9\lambda}(x)$
and $\psi_\lambda(x)$ are similar (though not identical) at all values
of $x,$ not only in their $T$ values, 
for $\lambda \simgt 3.$

Our reason for introducing 
the $\psi_\lambda$ distribution is primarily to make the
philosophical point that nongaussianity can be modeled 
phenomenologically,
and that as a result we
should be cautious in interpreting our findings as evidence of any
particular physical model.
Thus, although the cluster data point toward
a PDF with $T\approx 4,$ this does not imply that the quadratically
coupled inflationary model with $m\approx 40$ is correct. 
Indeed, any theory which invokes such a large number of identical
but independent scalar fields is suspect on Occam's Razor grounds alone.
More to the point, any PDF
with the indicated level of nongaussianity---such as the $\psi_\lambda$ distribution with $\lambda\simeq 3$---can can account for the cluster data.

Of course, even if
the $\psi_\lambda$ distribution proves able to describe the cluster
abundance data, this will not necessarily constitute evidence that
it is a good description of the PDF. The cluster data
test the positive tail of the PDF, not its shape near the peak,
which can only be probed via statistics sensitive to regions of
average density. The clustering of galaxies in the mildly
nonlinear regime, in which perturbation theory (PT) is valid, may
allow such a probe.  Frieman \& Gazta\~naga (1999) have applied
PT to the angular 3-point correlation function, and
have compared their predictions to the APM data. They are
able to rule out a strongly nongaussian PDF, the Peebles $\chi^2$
density field (cf. \S~\ref{sec:intro}), which has $T=16.3,$ 
via this approach. They did
not, however, consider the milder levels of nongaussianity ($T \approx 3$--4)
indicated by the cluster data for $\omegam\simeq 0.3.$ Inspection of
Figure~\ref{fig:phi_psi_1} shows that at such levels ($\lambda \sim 5$)
the PDF is close to Gaussian near its peak; it may be
quite difficult for the approach of Frieman \& Gazta\~naga to constrain
nongaussianity at this level. Another test of the PDF will be provided
by CMB anisotropy data. As mentioned at the outset of the paper,
analyses of the COBE data have unearthed indications of nongaussianity.
However, those results correspond to a comoving scale of \sm 1000\h1\ Mpc,
much larger than the \sm 10\h1\ Mpc cluster scales probed by clusters.
Future CMB measurements sensitive to $\simlt 10'$ scale
anisotropies will probe cluster mass scales directly. 
It is not yet clear, however,
that the signal-to-noise ratio per resolution element will be sufficient
to detect mild nongaussianity. It may well be  that for the
forseeable future the cluster data sets will provide the most
sensitive tests of PDF nongaussianity.

\subsection{What does it mean?}

If future tests confirm
that the initial density field is moderately
nongaussian, what would this tell us about the early universe? It was
argued above that this would not in and of itself lend credence to any
particular inflationary model. Rather, the basic 
significance of such a finding
would be that the
fundamental mechanism leading to real-space
Gaussianity of the initial fluctuations---their origin
as superpositions of numerous, statistically indpendent 
individual Fourier modes (equation~1)---is not fully operative.
This could be because either (i) only a small number of Fourier
modes contribute effectively to $\delta(\bfx),$ and the mode
amplitudes are themselves nongaussian; (ii)
the various Fourier modes are not statistically
independent; or (iii) some combination of (i) and (ii). 
It is the second of these effects which is responsible for nongaussianity
in Peebles' (1999a,b) model, 
but the first
cannot be excluded. Only as observational constraints on 
nongaussianity improve will be able to construct realistic models
of how it arises. This may lead us to a deeper
understanding of the fluctuations themselves.
Indeed, perhaps the greatest
lesson to be learned from this paper and the literature
cited herein is that we still know relatively little about the true
origin and character of the primordial density fluctuations.

\subsection{Reasons for Caution}

As seen in Figures~9 and~10, the PS-predicted comoving density
$N(\ge\!M,z)$ is a rapidly
decreasing function of mass for $M\simgt 10^{15}\,\h1\msun.$
Consequently, the predicted number of MS1054-like clusters
is extraordinarily sensitive to
the mass we assign to MS1054. We have selected this cluster
because three high-quality data sets, LK97, D98, and T99,
enable us to estimate its mass by three independent methods.
However, Figure~\ref{fig:ms1054_mass} shows that the velocity dispersion
mass estimate for MS1054 is about 50\% 
below the X-ray and lensing estimates. This discrepancy is within
the observational uncertainties and thus does not suggest the presence
of major systematic errors. On the other hand, if the velocity dispersion
mass estimate of T99 is correct, our conclusions
would be markedly changed. We would in that case find, using the
same criteria as above, that
Gaussian fluctuations are allowed for $\omegam\le 0.33$ 
as compared with $\omegam\leq 0.17,$ 
and that for $\omegam=0.45$ the 90\% constraint
is $\lambda\le 25$ ($T\ge 2$) as compared with
$\lambda \le 1$ ($T\ge 7$).
In short, 
our conclusion that substantial nongaussianity
is indicated by MS1054 is correct only
if the X-ray and lensing mass estimates are
closer to the truth than the dynamical one.

This extreme sensitivity to mass is both the strength and weakness
of the PS approach. It is a strength because it enables even one
high-redshift, high-mass cluster such as MS1054 to powerfully
constrain cosmology. It is a weakness because it means that even
modest random errors in mass estimation drastically affect our
quantitative conclusions. To best utilize the PS approach as a
cosmological probe, we will need much larger high-redshift
($z\simgt 0.6$) cluster samples than are presently available. 
Each sample cluster will need
to have its mass as accurately and robustly measured as MS1054.
Only then will the effect of mass errors be reduced, by $\sqrt{N}$ statistics,
to levels at which nongaussianity can be constrained (as a function of
$\omegam$) with high confidence. The EMSS cluster sample 
has been the major source of known intermediate and high redshift
clusters to date, but its sky coverage is too small at the faintest
X-ray flux levels to suffice for future work. Future distant cluster
samples will most likely be derived from deep optical surveys to which
automated cluster-finding algorithms (e.g., Postman \etal\ 1996; Kepner
\etal\ 1998) are applied, with cluster candidates followed up with
spectroscopy at 8--10 m class telescopes, X-ray satellite observations,
and ground- or space-based deep imaging, to obtain dynamical,
X-ray temperature, and weak lensing mass estimates respectively.

The second reason for caution is our reliance on the PS formalism, whose
validity at high masses is not yet confirmed. It has long been
known from N-body simulations 
that the PS formula overestimates the number of collapsed objects
at low masses, $M \ll M_*,$ where $M_*$ is the ``nonlinear mass'' at a given
epoch defined by $\sigma_M(M_*)=\delta_c(z).$ At masses
$M\simgt M_*,$ however, N-body studies have found
PS abundance predictions to be remarkably accurate
(see, e.g., Borgani \etal\ 1999 and references therein).
The nonlinear mass at the present time is $\sim 5\times 10^{13}\,\h1 \msun,$
so that rich clusters are safely above $M_*$ and thus
expected to be well-described by the PS formalism.
However, the PS formula has not been exhaustively tested in
the very high-mass ($M\simgt 100 M_*$) regime, for
the simple reason that most N-body simulations
contain very few objects in this mass range.

With the recent advent of extremely large
simulations, such tests have become possible.
In a set of simulations of cubic volumes
\sm 500\h1\ Mpc on a side, each containing hundreds
of rich clusters, Governato \etal\ (1998) were able
to test the accuracy of the PS abundance formula for
virial masses up to $\sim 3\times 10^{15}\,\h1\msun.$  
They found excellent agreement between the predicted
and observed number of clusters in simulations
of open, $\omegam=0.3$--0.4 universes. However,
in their $\omegam=1$ simulation with a
low present-day normalization ($\sigma_8\simlt 0.5$), they found
that the PS formula underpredicted the number
of clusters by a factor of \sm 3--10 for masses greater
than $\sim 10^{15}\,\h1\msun.$

The case for which Governato \etal\ found PS to be
inaccurate, $\omegam=1,$ is not one that is
relevant to this paper (we considered only $\omegam\leq 0.5$),
and for low-density universes (the case of interest here)
Governato \etal\ confirmed the PS abundance predictions.
Still, these results are cause for concern, because
if PS underpredicts abundances, we may be led to spurious
evidence for nongaussianty. On balance, then, while the present evidence
from N-body simulations favors continued
use of the PS formalism for cluster analysis, 
the Governato \etal\ findings suggest that
continued testing with larger simulations is needed.

\subsection{Conclusion}
We have used the generalized PS formalism to calculate the
likelihood that a cluster as massive as MS1054--03, at
a redshift of $z=0.83,$ would have been found in the 
EMSS sample. The calculations assumed a flat universe,
$\omegam+\omegal=1,$ and mass fluctuations $\sigma_M$
normalized to the observed abundance of low-redshift ($z\leq 0.1$)
clusters. The expected number of MS1054-like clusters then
depends on $\omegam$ and the deviations from Gaussianity
of the PDF, $P(\delta|M),$ on
cluster mass scales. We characterized
departures from Gaussianity
by assuming $P(\delta|M)=\sigma_M^{-1}\psi_\lambda(\delta/M),$
where $\psi_\lambda(x)$ is defined by equation~16.
The parameter $\lambda$ quantifies the degree of
nongaussianity; $\psi_\lambda$ approaches Gaussianity
for $\lambda \gg 1.$

Special attention has been given to the problem of estimating cluster
virial masses from galaxy velocity, X-ray temperature, and weak lensing
data. Such estimates are dependent on the assumed
density profile of the cluster, for which we have adopted the NFW form,
which has been shown in N-body simulations to be more realistic than
any pure power law. 
We elected to work with MS1054 alone because quality
X-ray temperature, galaxy velocity, and lensing data have
been obtained for it (D98, T99, LK97), making it a uniquely
well-studied high-redshift cluster at this time. The
X-ray and lensing mass estimates are in excellent agreement;
the dynamical mass is \sm 50\% smaller than the other two,
but this is within the observational errors.

If the initial density fluctuations
are Gaussian, it is improbable that a cluster of the mass,
redshift, and X-ray flux of MS1054 would be found in the EMSS if $\omegam\ge 0.2.$
For example, the chances of finding an MS1054-like cluster, for a Gaussian PDF,
in the EMSS search volume is less than
about one in fifty if $\omegam=0.25,$ and less than one in three hundred
if $\omegam=0.35,$ as shown in Figure~9. If the PDF is nongaussian,
however, the likelihood can be greatly enhanced.
To constrain the required amount of nongaussianity, we have determined the
value of $\lambda$ required for an MS1054-like cluster to have a one
in ten chance of being found in the EMSS. The results are shown
in Figure~\ref{fig:lambda_T} as a function of $\omegam.$ 
The maximum allowed value of $\lambda$ estimated using
this criterion decreases from $\sim 25$ for $\omegam=0.25$ to
$\sim 1$ for $\omegam=0.45.$ 
These results may also be
expressed in terms of the parameter $T$ 
(equation~20), which measures the likelihood of $\ge 3\sigma$
peaks in the density field relative to the Gaussian case. We find
$T\ge 2$ for $\omegam=0.25,$ increasing to $T\simgt 6$
for $\omegam\simgt 0.4.$ 
In short, for any value of $\omegam\simgt 0.25$
the initial density field must be significantly nongaussian, with
the extent of nongaussianity increasing rapidly with increasing $\omegam.$

Because our analysis has considered only the predicted
cluster {\em abundance,} it does not exclude Gaussian
fluctuations if we are willing to accept $\omegam < 0.2.$
However, the case for a nongaussian PDF irrespective
of $\omegam$ is strengthened if our results
are taken in conjunction with those of
RGS98 and KST99, who analyzed cluster correlation as well
as abundance data. In particular, 
KST99 found that nongaussianity was required, at the level
$T\simeq 2$--6, for {\em all\/} values of $\omegam$ less
than 0.5, and ruled out larger values of $\omegam.$ Our
estimate $T\ge 3$ for $\omegam\ge 0.3$ is consistent
with their results.
If one takes
the findings of RGS98, KST99, and this paper at face value,
the case for nongaussian initial density fluctuations is strong indeed.

We have, however, identified two potential weaknesses in our anlaysis,
and by extension
with any attempt to constrain cosmological parameters
from cluster abundance data. First, the predicted
abundances drop precipitously with increasing cluster virial mass at the
high masses ($\simgt 10^{15}\,\h1\msun$) and redshifts
($\simgt 0.5$) of interest. This sensitivity
translates into large errors in
the derived cosmological parameters for even modest ($\sim 30\%$)
errors in virial mass. 
This fundamental problem can only
be remedied by much larger catalogs
of high-redshift massive clusters; at present, MS1054 is
one of but a handful of such objects known. These clusters, once
identified, will need to be followed up with X-ray, galaxy velocity,
and weak lensing measurements to ensure reliable mass estimates.
Efforts are presently under way by this author and collaborators,
as well as other groups, to obtain such data sets. In 5--10 years
we will undoubtedly know much more than we do now about
the evolution of the cluster abundance at $z \sim 1.$

The second problem concerns the validity of the Press-Schechter formalism
in the high-mass regime. It has become conventional
wisdom in recent years that the PS abundance formula ``works much better
than it should,'' based as it is on a simple, spherical collapse model,
and as a result it has been widely and fruitfully used. However, 
if the PS formula underpredicts
the actual abundance of high-mass objects, as
at least one study suggests (Governato \etal\ 1998),
one would underestimate $\omegam$ (if Gaussian
fluctuations are assumed) or overestimate departures
from Gaussianity by applying the PS formula to cluster
abundance data. 
A challenge for theory and numerical simulations in the
coming years is to rigorously test the PS formula
at high masses and, if necessary, replace
it with a more accurate semianalytical framework. Such theoretical
groundwork will be crucially important if
the high-quality cluster data sets of
the coming decade are to be fully exploited.

\acknowledgements
The author gratefully acknowledges the support of
NSF grant AST-9617188, the Research Corporation,
and a Frederick Terman Fellowship from Stanford University.
The author thanks Nikhil Padmanabhan and
Puneet Batra for assistance in developing
the Press-Schechter computer codes, and 
Nikhil Padmanabhan, Puneet Batra, Keith Thompson,
and Sarah Church for useful discussions.  
Finally, the author would like to thank Joanne Cohn 
for bringing a number of papers on primordial
nongaussianity to his attention.

\vfill\eject
\appendix
\section{Approximating $\sigma_M(R)$}

\def\cali{{\cal I}}
In this appendix we obtain
an approximation for $\sigma_M(R)$ that is 
valid for CDM models. Using equations~5 and~6
in the main text, and assuming for simplicity $n=1,$
we may write the mass variance as follows:
\begin{equation}
\sigma_M^2(R) = \delta_H^2 \left(\frac{c}{H_0 R}\right)^4
\cali(\Gamma R)\,,
\label{eq:sigmr2}
\end{equation}
where 
\begin{equation}
\cali(\Gamma R) \equiv \int_0^\infty x^3 
W^2(x) T^2\left(\frac{x}{\Gamma R}\right)\,dx\,.
\label{eq:defcali}
\end{equation}
Equation~\ref{eq:sigmr2} further simplifies to
\begin{equation}
\ln\sigma_M(R)  = 16.012 + \ln\delta_H + 
0.5 \ln\cali(\Gamma R) - 2\ln R\,,
\label{eq:lncali}
\end{equation}
when $R$ is given in $\h1$ Mpc.

Equation~\ref{eq:lncali} shows that $\sigma_M(R)$
is fully determined, apart from a normalizing
constant involving $\delta_H,$ by 
$\ln\cali(\Gamma R).$
Since we fix the normalization by the cluster abundance
at low redshift, the $\delta_H$ term is
unimportant here, and a suitable expression for
$\cali(\Gamma R)$ is {\em all\/} we need to
evaluate $\sigma_M(R).$
We have obtained such an expression by numerically evaluating
$\cali(\Gamma R)$ for a range $\Gamma R$
and fitting the results to
a low-order polynomial of the form
\begin{equation}
\ln\cali(\Gamma R) = \sum_{n=0}^m a_n\left[\ln(\Gamma R)\right]^n\,.
\label{eq:fit}
\end{equation}
Choosing $m=5$ produced a fit
with better than $0.1\%$ accuracy
for $1 \leq \Gamma R \leq 8,$ fully covering the range
of expected values. 
In Table~\ref{tab:fitcoff} we give the coefficients 
yielding the best fifth-order fit.
Substituting this polynomial expression
into equation~\ref{eq:lncali} yields $\sigma_M$
up to a normalizing coefficient. The expression for
the logarithmic derivative of $\sigma_M$ needed
for the PS abundance formula is then simply
\begin{equation}
\frac{d\ln\sigma_M}{d\ln R} = \frac{1}{2}\sum_{n=1}^m
n a_n \left[\ln(\Gamma R)\right]^{n-1} \, - \, 2\,.
\label{eq:dlnsig}
\end{equation}
\begin{table*}
\begin{minipage}{180mm}
\caption[ ]{\centerline{{\footnotesize
Fit Coefficients for Computing $\ln\cali(\Gamma R)$}}}
\centerline{
\begin{tabular}{c c c c c c }
\hline\hline\multicolumn{6}{c}{} \\
$a_0$ & $a_1$ & $a_2$ &
$a_3$ & $a_4$ & $a_5$  \\ \hline
$-4.7783$ &
$ 2.7251$ &
$-0.1811$ &
$-0.0232$ &
$-0.0053$ &
$ 0.0014$ \\ \hline\hline
\label{tab:fitcoff}
\end{tabular}
}
\smallskip

Notes: Coefficients in the expansion
defined by equation~\ref{eq:fit}.

\end{minipage}
\end{table*}

\end{document}